\documentclass[twocolumn,showpacs,amsmath,amssymb]{revtex4}
\usepackage[cp866]{inputenc}
\usepackage[english]{babel}
\usepackage{graphicx}
\usepackage{bm}
\usepackage{epsf}
\usepackage{amssymb,amsmath}
\begin{document}

\title
{Electron-muon heat conduction in neutron star cores via
the exchange of transverse plasmons}

\author{P.~S.~Shternin 
        and D.~G.~Yakovlev}
\affiliation{
        Ioffe Physico-Technical Institute,
        Politekhnicheskaya 26, 194021 Saint-Petersburg, Russia}
\date{\today}

\newcommand{\sgn}{{\rm sign}}

\begin{abstract}
We calculate the thermal conductivity of electrons
and muons $\kappa_{e\mu}$ produced 
owing to electromagnetic interactions 
of charged particles in neutron star
cores and show that these interactions are dominated by the
exchange of transverse plasmons (via
the Landau damping of
these plasmons  in nonsuperconducting matter and 
via a specific plasma screening in the presence of
proton superconductivity). For normal protons, the Landau
damping strongly reduces $\kappa_{e\mu}$ and makes it
temperature independent.
Proton superconductivity suppresses the reduction and
restores the Fermi-liquid behavior $\kappa_{e\mu}\propto T^{-1}$.
Comparing with the thermal conductivity of neutrons $\kappa_n$,
we obtain $\kappa_{e\mu}\gtrsim \kappa_n$ for 
$T \gtrsim 2 \times 10^9$~K in normal matter and 
for any $T$ in superconducting matter 
with proton critical temperatures $T_{cp} \gtrsim 3 \times 10^9$~K. 
The results are described by simple analytic formulae.
\end{abstract}


\maketitle

\section{Introduction}
\label{introduc}

Neutron stars are compact stars of mass $ \sim 1.4\,M_\odot$
and radius $\sim 10-15$ km, where $M_\odot$ is the solar mass.
A neutron star is thought to consist of a thin crust
($\sim 1$\% by mass) and a bulky core \cite{st83,lp01,haensel03,lp04}.
The crust is composed of a nonuniform matter containing
atomic nuclei. The core extends from the layer of
the density $\rho \approx 0.5\,\rho_0$
to the stellar center [$\rho \sim (10-20)\,\rho_0$],
where $\rho_0 \approx 2.8 \times 10^{14}$~g~cm$^{-3}$
is the mass density of saturated symmetric nuclear matter.
It is widely thought that the core is built of neutron-rich,
strongly degenerate, uniform nuclear
matter of supernuclear density. This makes neutron stars
unique natural laboratories of such matter. The composition
and equation of state (EOS) of this matter are largely unknown
but will hopefully be strongly constrained soon by comparing theoretical
models of neutron star structure and evolution with
observations.

We will consider the thermal conductivity of neutron star cores.  The
thermal conductivity of neutron stars is needed for modeling the
thermal structure and evolution of these stars and related
phenomena \cite{yp04,pgw06}. In particular, it is important for
cooling of isolated neutron stars and for thermal structure of
transiently accreting neutron stars (in soft X-ray transients).
Accreting stars can be heated by pycnonuclear reactions when
accreted matter sinks into the deep neutron star crust under the
weight of newly accreted material \cite{hz90,hz03,bbr98}. Thermal
conduction spreads this heat over the star -- to the surface
(resulting in surface photon emission) and to the core (producing
neutrino emission from the core). In this way the observable
surface thermal emission depends on the neutrino luminosity of the
star, and, hence, on its internal structure. The thermal
conductivity of neutron star cores is especially important for
studying cooling of young neutron stars (of age $t \lesssim 100$
years), where the internal thermal relaxation is not yet achieved
(e.g., Refs.\
\cite{lattimeretal94,gyp01}).
The thermal conductivity can also
regulate thermal relaxation of a neutron star in
response to pulsar glitches
(see, e.g.,
\cite{ll02}, and references therein).

We will focus on models of neutron star cores
composed of neutrons ($n$), with an admixture of protons ($p$),
electrons ($e$), and muons ($\mu$). Our
results are easily generalized for matter containing
hyperons. In all these cases, the thermal conductivity $\kappa$ can be
subdivided into two relatively independent parts \cite{fi79,fi81},
%
\begin{equation}
    \kappa \approx \kappa_b + \kappa_{e\mu},
\label{kappa-total}
\end{equation}
where $\kappa_b$ is the conductivity
of neutrons (and possibly other baryons) mediated by
strong-interaction collisions of heat-conducting particles
(see, e.g., 
\cite{bhy01},
and references therein), while
$\kappa_{e\mu}$ is the conductivity
of electrons and muons mediated by electromagnetic interactions.
The conductivity $\kappa_{e\mu}$ has been calculated
in several papers, particularly,
in Refs.\ \cite{fi76,fi79,gy95}.

However, all previous calculations of
$\kappa_{e\mu}$ in neutron star cores have
neglected an important effect of the Landau damping 
of electromagnetic interactions 
owing to the exchange of transverse plasmons. In the context of
transport properties of dense matter
this effect was studied by Heiselberg and Pethick
\cite{hp93}   
for a degenerate quark plasma.
Recently we have included the effect of the Landau
damping into the thermal conductivity
of strongly degenerate electrons determined by electron-electron collisions
with the electron plasma screening
\cite{sy06}.
Similar effects have
also been studied by Jaikumar {\it et al}.\
\cite{jaikumaretal05} 
for neutrino bremsstrahlung radiation via electron-electron
collisions in neutron star crusts and cores.
Here we reconsider
$\kappa_{e\mu}$ in a neutron star core
including the effects of the Landau
damping and of the specific plasma screening induced
by proton superconductivity.

\section{Electron-muon thermal conductivity in normal matter}
\label{formalism}

The electrons and muons in a neutron star core constitute
strongly degenerate almost ideal Fermi gases \cite{st83}. The electrons are
ultrarelativistic. The muons can be absent in the outermost
part of the core; they are nonrelativistic just after their
creation threshold and  become relativistic
at higher densities. The neutrons and protons constitute
strongly interacting Fermi liquids. 
The neutrons are mainly nonrelativistic (although they become
mildly relativistic near the center of a massive neutron star);
the less abundant protons are typically nonrelativistic.
The neutrons and protons can be in superfluid states;
superfluidity of protons means their superconductivity. 
In this section we consider nonsuperfluid (normal) matter.
We analyze the effects of proton
superconductivity in Sec.\ \ref{sect-superfluid}.
We closely follow the derivation
of $\kappa_{e\mu}$ by Gnedin and Yakovlev \cite{gy95}
(but introduce the Landau damping); thus we omit
technical details.

The thermal conductivity $\kappa_{e\mu}$ is calculated from
a coupled system of linearized Boltzmann equations for
the electron and muon distribution functions $\widetilde{f}_e$ and
$\widetilde{f}_\mu$, which we denote collectively
by $\widetilde{f}_c$, with $c=e$ or  $\mu$.
These distributions slightly deviate from the equilibrium Fermi-Dirac
distributions $f_c$ owing to the presence of a weak
temperature gradient $\bm{\nabla} T$,
\begin{eqnarray}
    \widetilde{f}_c&=&f_c -\Phi_c\,    
    {\partial f_c  \over \partial \varepsilon_c},
\nonumber \\    
    f_c&=& \left\{ \exp \left( {\varepsilon_c - \mu_c \over
    k_B T} \right) +1 \right\}^{-1},
\label{distributions}
\end{eqnarray}
where $\varepsilon_c$ is the particle energy, $\mu_c$ is its
chemical potential (the rest-mass term included),
$T$ is the temperature, $k_B$ the Boltzmann
constant, and $\Phi_c$  measures a deviation from equilibrium.
For calculating $\kappa_{e\mu}$,
the electrons and muons are treated as the only heat carriers
which experience electromagnetic interactions between themselves
and with any charged baryons in dense matter (with protons, in our
case). The charged baryons are assumed to obey
equilibrium Fermi-Dirac distributions
(see, e.g., Ref.\ \cite{gy95}).
Nonequilibrium parts of the electron and muon
distributions are found using the standard variational approach with
the simplest trial function,
\begin{equation}
     \Phi_c= -\tau_c \, (\varepsilon_c-\mu_c)\, \bm{v}_c \cdot
     \bm{\nabla} T/T,
\label{Phi}
\end{equation}
where $\bm{v}_c$ is the velocity of particles $c$ and $\tau_c$ is their
effective 
relaxation time.

The resulting electron and muon thermal conductivity can be
written \cite{gy95} in a familiar form
\begin{eqnarray}
  &&   \kappa_{e\mu}=\kappa_e+\kappa_\mu, 
 \nonumber \\
  &&  
 \kappa_e={\pi^2 k_B^2 T n_e \tau_e \over 3m_e^*}, \quad
     \kappa_\mu={\pi^2 k_B^2 T n_\mu \tau_\mu \over 3m_\mu^*},
\label{kappa}
\end{eqnarray}
where $\kappa_e$ and $\kappa_\mu$ are
the partial thermal conductivities of electrons and muons, respectively;
$n_e$ and $n_\mu$ are number densities of these particles;
$m_e^*=\mu_e/c^2$ and $m_\mu^*=\mu_\mu/c^2$ are their
effective masses.
In neutron star cores at beta equilibrium
one has $\mu_e=\mu_\mu$ and
$m_e^*=m_\mu^*$ \cite{st83}.
Note that the thermal conductivities $\kappa_e$
and $\kappa_\mu$ presented by Gnedin and Yakovlev \cite{gy95}
[their Eqs.\ (61) and (65)] contain also an additional
factor $C\approx 1.2$ 
which brings the employed simplest
variational solution
closer to the exact solution.
We do not introduce a similar correction
here. Its calculation is more complicated than in Ref.\ \cite{gy95},
because now we include the Landau damping of transverse
plasmons. The Landau damping introduces non-Fermi-liquid behavior into
the thermal conductivity (Sec.\ \ref{sect-coll-freq}), while
the standard technique for calculating the corrections
is developed for traditional Fermi liquids.
In Sec.\ \ref{sect-exact} we argue that the Landau
damping of transverse plasmons reduces the difference between
the exact and simplest variational solutions. Thus,
neglecting the correction factor $C$ in the
expression for $\kappa_{e\mu}$ can introduce uncertainties
$\lesssim$20\% which are quite acceptable for
the thermal conduction problem in neutron star cores.

The effective electron and muon
relaxation times can be written as \cite{gy95}
\begin{equation}
   \tau_e =
   {\nu_\mu -\nu'_{e\mu} \over \nu_e \nu_\mu - \nu'_{e\mu}\nu'_{\mu e}},
   \quad
   \tau_\mu =
   {\nu_e -\nu'_{\mu e} \over \nu_e \nu_\mu - \nu'_{e\mu}\nu'_{\mu e}},
\label{taus}
\end{equation}
where
\begin{eqnarray}
     \nu_e &=& \sum_i \nu_{ei}=\nu_{ee}+\nu_{e \mu}+\nu_{ep}, 
\nonumber \\     
     \nu_\mu&=& \sum_i \nu_{\mu i}=\nu_{\mu \mu}+ \nu_{\mu e}+ \nu_{\mu p}
\label{nus}
\end{eqnarray}
are the total effective collision frequencies of electrons
and muons with all charged particles $i$ (which are electrons,
muons and protons, in our case); $\nu_{ei}$ and $\nu_{\mu i}$
are familiar partial effective collision frequencies (derived in a
standard manner from respective linearized collision integrals),
while $\nu'_{e \mu}$ and $\nu'_{\mu e}$ are two additional
effective collision frequencies \cite{gy95} which couple
heat transport of the electrons and muons.
All these  collision frequencies can be
expressed as multi-dimensional integrals over
momenta of colliding particles.
Since all particles are
strongly degenerate, their momenta can be placed at
appropriate Fermi surfaces whenever possible.
In this way, for collisions of nonidentical
particles 1 and 2 we obtain
\begin{eqnarray}
  \nu_{12}&= &{3 \over T^3 v_{F1} p^2_{F1}}\, (S_{11}-S_{11'}),
  \nonumber \\
  \nu'_{12}&= &{3 \over T^3 v_{F1} p^2_{F1}}\, (S_{12}-S_{12'}),
\label{nunu}
\end{eqnarray}
while for collisions between identical heat carriers
we have $\nu_{cc}=\nu_{12}+\nu'_{12}$.
Here, $v_{F1}$ and $p_{F1}$ are, respectively, the Fermi velocity and
the Fermi momentum of particles 1;
\begin{eqnarray}
    S_{\alpha \beta} & = & (2 \pi)^{-12} \, \int
    {\rm d}^3p_1\,{\rm d}^3p_1'\,{\rm d}^3p_2\,{\rm d}^3p_2'\,
\nonumber \\    
    && \times W(1,2|1',2')\,f_1 f_2 (1-f_1')(1-f_2')
\nonumber \\
    & & \times (\bm{v}_\alpha \cdot \bm{v}_\beta)\,
        (\varepsilon_\alpha - \mu_\alpha)\,
    (\varepsilon_\beta - \mu_\beta)
\label{S12}
\end{eqnarray}
is the normalized collision frequency;
\begin{eqnarray}
 W(1,2|1',2')&=&4\,(2 \pi)^4 \, \delta(\varepsilon_1+
 \varepsilon_2-\varepsilon'_1 - \varepsilon'_2)
\nonumber \\ 
&& \times \delta( \bm{p}_1+\bm{p}_2-\bm{p}'_1-\bm{p}'_2)\,|M_{12}|^2
\label{W12}
\end{eqnarray}
is the differential transition probability for a scattering
process $\bm{p}_1 \bm{p}_2 \to \bm{p}'_1 \bm{p}'_2$;
$|M_{12}|^2$ is the squared matrix element summed over particle
spin states (it includes also the symmetry factor
to avoid double counting of the same collisions events);
primes refer to particles after the collision.
In Eqs.\ (\ref{nunu})--(\ref{W12}) we
use the system of units in which $c=k_{\rm B}=\hbar=1$.
The same system will be used below unless the contrary
is indicated.

In the absence of muons we have
\begin{equation}
   \kappa_{e\mu}=\kappa_e, \quad  \tau_e^{-1}=\nu_e= \nu_{ee}+\nu_{ep}.
\label{nomuons}
\end{equation}
%

\subsection{Dynamical screening of electromagnetic interaction}
\label{screening}

The physics of the dynamical plasma screening
is thoroughly analyzed by Heiselberg and Pethick \cite{hp93}.
These authors consider quark-quark collisions in a quark plasma through
one-gluon exchange in the weak-coupling limit. Such collisions
are similar to electromagnetic scattering of
charged particles in an ordinary plasma.
Electromagnetic interactions of
muons and electrons in neutron star cores are
usually accompanied by small momentum
and energy transfers which greatly simplifies the theory.
The squared matrix element for small energy
transfers in a collision
$\bm{p}_1 \bm{p}_2 \to \bm{p}'_1 \bm{p}'_2$
of nonidentical particles 1 and 2 is
\begin{equation}
      |M_{12}|^2 \propto \left| \frac{J_{1'1}^{(0)}
      J_{2'2}^{(0)}}{q^2+\Pi_l}-
      \frac{\bm{J}_{t1'1} \bm{\cdot} \bm{J}_{t2'2}}{q^2-\omega^2+\Pi_t}
      \right|^2,
\label{matelement}
\end{equation}
where $\hbar \bm{q}=\bm{p}_1'-\bm{p}_1$ is
a momentum transfer, $\hbar \omega =\varepsilon_1'-\varepsilon_1$
is an energy transfer (in standard physical units),
$J_i^{(\nu)}=(J_{i'i}^{(0)},\bm{J}_{i'i})=
(\bar{u}_{i}'\gamma^\nu u_i)$ is the transition 4-current
($\nu$=0,\ldots,3), $\gamma^\nu$ is a Dirac matrix,
$u_i$ is a normalized bispinor ($\bar{u}_i u_i=2 m_i$),
$\bar{u}_i$ is a Dirac conjugate (see, e.g.,
Berestetski{\u\i}, Lifshitz  and Pitaevskii \cite{QED});
$\bm{J}_{ti'i}$ is the component of $\bm{J}_{i'i}$
transverse to $\bm{q}$.
The longitudinal component of $\bm{J}_{i'i}$ (parallel
to $\bm{q}$) is related to the time-like (charge density)
component $J^{(0)}_{i'i}$ via charge conservation. It is
excluded from Eq.~(\ref{matelement}) with the aid of
the continuity equation as explained by Heiselberg and
Pethick \cite{hp93}. For collisions of identical particles
($ee$ and $\mu \mu$ in our case), the matrix
element contains two parts, $M_{12}=M^{(1)}_{12}+M^{(2)}_{12}$,
which correspond to two channels, ($1\to1';2\to2'$) and
($1\to2';1'\to2$). However, in the small-momentum-transfer
approximation,
the interference term is small,
both channels give equal contributions,
and the relationship (\ref{matelement}) is not violated.

Equation (\ref{matelement}) contains
the polarization functions $\Pi_l$ and $\Pi_t$ which
depend on $\omega$ and $q$ and
describe plasma screening of
interparticle interaction by longitudinal
and transverse plasma perturbations (plasmons), respectively.
In the random phase approximation (RPA), these functions are the sums of
terms for all charged particles $i$
(electrons, muons and protons).
In the classical limit ($q \ll p_{Fi}$,
$\omega \ll v_{Fi} p_{Fi}$) one has
(e.g., Alexandrov, Bogdankevich and Rukhadze \cite{abr84})
\begin{eqnarray}
     \Pi_l&= &\sum_i {4 \alpha p_{Fi}^2 \over \pi v_{Fi}}\,
     \chi_l(x_i), 
\nonumber \\     
     \Pi_t &= &\sum_i {4 \alpha p_{Fi}^2 v_{Fi}\over \pi }\,
     \chi_t(x_i),
\label{polariz}
\end{eqnarray}
where $x_i=\omega/(q v_{Fi})$,
$\alpha=e^2/\hbar c\approx 1/137$ is the fine
structure constant,
and
\begin{eqnarray}
     \chi_l(x) &=&  1- {x \over 2}\,
    \ln \left( x+1 \over x-1 \right) ,
\nonumber \\
    \chi_t(x) &=&  {x^2 \over 2} +
    { x(1-x^2) \over 4}\, \ln \left( x+1 \over x-1 \right) .
\label{chi}
\end{eqnarray}

For typical conditions of very strong degeneracy in neutron star cores,
it is sufficient to use the expressions for $\Pi_l$ and $\Pi_t$
in the limit of $q \ll p_{Fi}$ and $x_i \ll 1$,
in which $\chi_l(x) \approx 1$ and $\chi_t(x) \approx i \pi x/4$.
In this limit,
\begin{eqnarray}
   \Pi_l&=&\sum_i \frac{3\omega_{i}^2}{v_{Fi}^2} = q_l^2,
\label{Pil}   \\
   \Pi_t&=& i \, \frac{\pi}{4} \, \frac{\omega}{q}\sum_i
   \frac{3\omega_{i}^2}{v_{Fi}}=i \,
   \frac{\pi}{4} \, \frac{\omega}{q} \, q_t^{2},
\label{Pit}
\end{eqnarray}
where $\omega_i=(4 \pi e^2 n_i/m_i^*)^{1/2}$
is the plasma frequency of particles $i$, and
$m_i^*$ is the effective particle mass at the
appropriate Fermi surface. We have already defined
$m_e^*$ and $m_\mu^*$ for electrons and muons
which form almost ideal Fermi gases. Their effective
masses differ from the bare masses owing to relativistic effects.
As for protons ($i=p$), which are nonrelativistic in neutron star
cores, their effective mass $m_p^*$ differs from the bare proton mass
due to strong interactions with surrounding nucleons.
In the approximation (\ref{Pil}) and (\ref{Pit})
one can also neglect $\omega^2$ in the denominator of the
second term in (\ref{matelement}).
Equations (\ref{chi}) are strictly valid for Fermi gases, which is
a good approximation for the electrons and muons, 
but the protons constitute a Fermi liquid.  
Fortunately, the
asymptotic expressions (\ref{Pil}) and (\ref{Pit}), sufficient for
our calculations, remain valid for the Fermi liquid.
This is because a kinetic equation 
for quasiparticles in the Fermi liquid at low $\omega$
is similar to a familiar
kinetic equation for  
Fermi gases (see, e.g., Ref.\ \cite{lp80}).

In Eqs.\ (\ref{Pil}) and (\ref{Pit})
we have introduced $q_l$ and $q_t$ [cm$^{-1}$] defined as
(in standard physical units)
\begin{eqnarray}
   q_l^2&=& \frac{4\alpha}{\pi} \sum_i
   {c m_i^* p_{Fi}\over \hbar^2},
\label{ql}   \\
   q_t^2&=&\frac{4\alpha}{\pi} \sum_i
   {m_i^* p_{Fi} v_{Fi}\over \hbar^2}
   =\frac{4\alpha}{\pi} \sum_i {p_{Fi}^2 \over \hbar^2}.
\label{qt}
\end{eqnarray}
Generally, we have $q_t \leq q_l$.
If all charged particles were ultrarelativistic then $q_t \to q_l$.

According to Eqs.\ (\ref{Pil}) and (\ref{Pit}), the screening
functions for longitudinal and transverse plasmons are
different. The screening via longitudinal plasmons
(\ref{Pil}) is described by a real (nondissipative) polarization
function $\Pi_l=q_l^2$ which is equivalent to a Debye-like
(Thomas-Fermi) plasma
screening with the screening length $1/q_l$. The screening of
transverse currents is described by a purely imaginary
(dissipative) function (\ref{Pit}), corresponding to the Landau
damping via collisionless absorption of 
transverse plasmons by plasma particles.
Calculating the electron and muon thermal conductivity in neutron
star cores, the authors of Refs.\ \cite{fi79,gy95}
have neglected the Landau damping and erroneously used the
same static Debye-like screening, $\Pi_t=q_l^2$, for transverse
currents as for longitudinal ones. We will show that this
approximation strongly overestimates the thermal conductivity of
electrons and muons.

The squared matrix element (\ref{matelement})
for collisions between heat carriers
$c$ and scatterers $i$ can be written as
\begin{eqnarray}
   |M_{ci}|^2 = \frac{16\pi^2\alpha^2}{m_c^{*2}m_i^{*2}}\, \varphi,
\quad
   \varphi= \varphi_\parallel+
   \varphi_\perp+\varphi_{\perp\parallel},
\label{f}
\end{eqnarray}
where $\varphi$ is a dimensionless function
determined by the exchange of longitudinal
plasmons ($\varphi_\parallel$), transverse plasmons
($\varphi_\perp$), and by both interactions
($\varphi_{\parallel \perp}$). The leading terms in 
the
weak screening approximation are
\begin{eqnarray}
    \varphi_{\parallel}&=&\frac{m_c^{*2}m_i^{*2}}{(q^2+q_l^2)^2},
\label{fpar}\\
    \varphi_{\perp}&=&
    \frac{p_{Fc}^2 p_{Fi}^2 q^2 \cos^2\phi}{q^6+\Lambda^6},
\label{fper}\\
\varphi_{\perp\parallel}&=&
    -\frac{2m_c^*m^*_i
    p_{Fi} p_{Fc} q^4\cos\phi}{(q^2+q_l^2)(q^6+\Lambda^6)},
\label{fcomb}
\end{eqnarray}
where $\Lambda= (\pi\omega q_t^2/4)^{1/3}$ and $\phi$ is the
angle between $\bm{p}_1$ and $\bm{p}_2$. For collisions of
identical particles ($i=c$), the function $\varphi$ is generally
different but the difference is negligible in the leading-order
weak screening approximation.

\subsection{Calculation of collision frequencies}
\label{sect-coll-freq}

Calculating the integrals (\ref{S12})
for a strongly degenerate matter,
we can use the standard angular-energy decomposition
(e.g., Ref.\ \cite{st83}).
It consists in placing momenta of colliding particles
at respective Fermi surfaces, whenever possible, and in
decoupling energy and
angular integrations using ${\rm d}^3p=m^*p_F
\,{\rm d}\varepsilon\,{\rm d}\Omega$.
The squared matrix element in (\ref{W12})
depends only on relative orientations of momenta of colliding particles
which greatly simplifies angular integrations.
Performing similar transformations as in
Gnedin and Yakovlev \cite{gy95} we can reduce Eqs.\ (\ref{nunu}) and
(\ref{S12}) for the collision frequencies of nonidentical
particles to
\begin{eqnarray}
   \nu_{ci}&=&\frac{12\alpha^2}{\pi^4 T^3 p_{Fc} m_c^*}
   (J_{ci1}-J_{ci1'}),
\label{nuJ}\\
   \nu'_{ci}&=&\frac{12\alpha^2 p_{Fi}}{\pi^4 T^3 p_{Fc}^2 m_i^*}
   (J_{ci2}-J_{ci2'}).
\label{nuJc}
\end{eqnarray}
Here,
\begin{eqnarray}
    J_{cik}&=&\int\int\int\int
    {\rm d}\varepsilon_1 \,
    {\rm d}\varepsilon_1'\,
    {\rm d}\varepsilon_2 \,
    {\rm d}\varepsilon_2'\, f_1f_2   
    (1-f_1')(1-f_2')
   \nonumber \\
    && \times (\varepsilon_1-\mu_1)(\varepsilon_k-\mu_k)I_{\Omega k}(\omega)
\end{eqnarray}
is an energy integral and
$I_{\Omega k}(\omega)$ is a remaining angular integral, which
depends on a collision energy transfer $\omega$ owing to the
dynamical screening in Eq.~(\ref{Pit}).
The effective collision frequency $\nu_{cc}$ 
between identical heat carriers
can be calculated as $\nu_{cc}=\nu_{ci}+\nu'_{ci}$ with $i=c$.

In the weak screening approximation we obtain
\begin{eqnarray}
    I_{\Omega 1}&=&I_{\Omega 1'}= \int_0^{q_m}
    {\rm d}q \int_0^{\pi} {\rm d}\phi \, \varphi,
\label{IOmega1}\\
    I_{\Omega 2}&=&I_{\Omega 2'}=\int_0^{q_m} {\rm d}q \int_0^{\pi}
    {\rm d} \phi
    \, \varphi \cos \phi,
\label{IOmega2}
\end{eqnarray}
with $q_m=\min\{2p_{Fi},2p_{Fc}\}$. The integration
over $\phi$ is then trivial. From Eqs.\
(\ref{fpar})--(\ref{fcomb}) we see that $I_{\Omega 1}$ consists
of two terms associated with the longitudinal and
transverse plasmon exchange, respectively,
while $I_{\Omega 2}$ contains only
the mixed term,
\begin{equation}
   I_{\Omega 1}= I^{\perp}_{\Omega 1}+I^{\parallel}_{\Omega 1}, \quad
   I_{\Omega 2}= I^{\perp\parallel}_{\Omega 2}.
\end{equation}
This means, that $\nu_{ci}$ is determined by the contributions
from longitudinal and transverse screening, but $\nu'_{ci}$ is
determined by the mixed term.

We have calculated the angular integrals in the weak screening
approximation, retaining the leading terms in series expansions 
over $q_l$ or $\Lambda$. The result is
\begin{eqnarray}
    I^{\perp}_{\Omega 1} &=&\frac{\pi}{3}
    \frac{p_{Fi}^2 p_{Fc}^2}{|\omega|}\frac{1}{q_{t}^2},
    \\
    I^{\parallel}_{\Omega 1} &=& \frac{\pi^2m_c^{*2}m_i^{*2}}{4q_l^3},
    \\
    I^{\perp\parallel}_{\Omega 2}
    &=&-\frac{\pi^2}{3}\left(\frac{4}{\pi}\right)^{1/3}
    \frac{p_{Fi} p_{Fc}}{q_l^2}
    \frac{m_i^*m_c^*}{q_t^{2/3}|\omega|^{1/3}}.
\end{eqnarray}

Using the standard technique (e.g., \cite{gy95}) we
can reduce the energy integration to
\begin{eqnarray}
   J_{ci1}-J_{ci1'}&=& \int_0^\infty
   \frac{I_{\Omega 1}(\omega)\, \omega^3\,
   \exp(-\omega/T)\,{\rm d}\omega }
   {\left[1-\exp(-\omega/T)\right]^2},
\label{Jci1}\\
   J_{ci2}-J_{ci2'}&=&
   -\int_0^\infty \frac{I_{\Omega 2}(\omega)\,
    \omega^3 \, \exp(-\omega/T)\,{\rm d}\omega}
    {\left[1-\exp(-\omega/T)\right]^2}.
\label{Jci2}
\end{eqnarray}

The final $\omega$ integration
gives analytical expressions for the collision frequencies
(in standard physical units)
\begin{eqnarray}
     \nu_{ci}&=&\nu_{ci}^\perp+\nu_{ci}^\parallel,
\label{nuci}\\
     \nu^{\perp}_{ci}&=&\frac{24\zeta(3)}{\pi^3}\,
     \frac{\alpha^2 k_B T}{\hbar c m_c^*}
     \,\frac{p_{Fi}^2p_{Fc}}{\hbar^2 q_t^2},
\label{nuciperp}\\
     \nu_{ci}^{\parallel}&=&\frac{4\pi^2}{5}\,
     \frac{\alpha^2 k_B^2 T^2}{\hbar c^2 m_c^*}\,
     \frac{c^4 m_c^{*2}m_i^{*2}}{p_{Fc} \hbar^3 q_l^3},
\label{nucipar}\\
     \nu'_{ci}&=&\frac{12 \xi}{\pi^3} \,
     \frac{\alpha^2 (k_B T)^{5/3}}{\hbar (\hbar c q_t)^{2/3}} \,
     \frac{c m_c^*p_{Fi}^2}{\hbar^2 p_{Fc} q_l^2},
\label{nucicross} \\
     \xi &=& {(2 \pi)^{2/3} \over 3}\,
     \Gamma\left(14 \over 3\right)\,
     \zeta\left(11 \over 3\right)
     \approx 18.52,
\label{xi}
\end{eqnarray}
where $\zeta(z)$ is the Riemann zeta function, 
and $\Gamma(z)$ is the gamma function.
Equations (\ref{kappa})--(\ref{nus}) and
(\ref{nuci})--(\ref{nucicross}) fully
determine the electron and muon thermal
conductivity of a nonsuperconducting matter calculated in the
weak screening approximation.

Owing to a strong degeneracy, for typical conditions in
neutron star cores
\begin{equation}
    \nu_{ci}^\parallel\ll \nu_{ci}'\ll \nu_{ci}^\perp.
\label{nuinequality}
\end{equation}
Because $\nu'_{e \mu}$ and $\nu'_{\mu e}$ are negligible,
electron and muon heat transports ($c= e$ and $\mu$) become formally
decoupled. In this approximation from Eq.\ (\ref{taus}) we obtain
\begin{equation}
   {1 \over \tau_c}= \nu_c^\perp =
   \sum_i \nu_{ci}^\perp = \frac{6\zeta(3)}{\pi^2} \,
   \frac{\alpha k_B Tp_{Fc}}{\hbar c m_c^*}.
\end{equation}
Here the electron and muon thermal relaxation times $\tau_e$ and
$\tau_\mu$ are mostly determined by the exchange of transverse
plasmons (by the Landau damping) and are expressed only through
the parameters of heat carriers ($e$ or $\mu$).

The thermal conductivity of the electrons
or muons becomes temperature independent,
\begin{equation}
\label{E:kappa_normal}
    \kappa_c=\kappa_c^\perp=\frac{\pi^2}{54\zeta(3)}\,
    \frac{k_B c p_{Fc}^2}{\hbar^2 \alpha},
\end{equation}
being solely determined by the Fermi momentum $p_{Fc}$
of heat carriers.

The remarkable simplicity of the thermal conductivity
of strongly degenerate particles in this regime, where
electromagnetic interactions are dominated by
the Landau damping of transverse plasmons,
was pointed out by Pethick and
Heiselberg \cite{hp93} for a plasma of massless (ultrarelativistic)
quarks. The authors stressed that the
temperature dependence of this thermal conductivity
strongly deviates from the dependence
$\kappa \propto T^{-1}$ in an ordinary Fermi liquid,
because the plasma screening
due to the exchange of transverse plasmons depends on
energy transfer $\hbar \omega$ in collision events.
Were particle collisions dominated by the exchange
of longitudinal plasmons ($\nu=\nu^\parallel$), the
thermal conductivity would show the traditional
Fermi-liquid behavior (as was obtained
in Refs.\ \cite{fi76,fi79,gy95},
whose authors
erroneously assumed the Debye-like plasma screening, the same
as determined
by the exchange of longitudinal plasmons).

Recently we \cite{sy06}
have calculated
$\nu_{ee}$, taking proper account of the electron plasma screening via
the exchange of longitudinal and transverse plasmons, 
for degenerate electrons of
any degree of relativity. We have used more general electron
polarization functions given by Eqs.\ (\ref{polariz}) and (\ref{chi})
[rather than by the asymptotic limits (\ref{Pil}) and
(\ref{Pit}) employed in the present paper].
We have applied our results for studying the electron
thermal conductivity in a neutron star crust, but 
after adding the plasma screening by muons and protons
[see Eqs.\ (\ref{ql}) and (\ref{qt})]
they
become valid for nonsuperconducting
neutron star cores. 
Replacing in $\nu_{ee}$ electrons
by muons we can obtain the expression for $\nu_{\mu \mu}$.

Our present results include electromagnetic interactions of
electrons, muons and protons 
(not only $\nu_{ee}$ or $\nu_{\mu \mu}$ which could be extracted
from the results of Ref.\ \cite{sy06}); 
they are valid for degenerate particles of any degree
of relativity. Because the exchange of transverse plasmons
is efficient for relativistic particles, it becomes
unimportant if all charged particles are 
nonrelativistic. The presence of ultrarelativistic electrons
in neutron star cores ensures the importance of the Landau
damping there.

\subsection{Comparison with exact solution}
\label{sect-exact}

So far we have used a simplest variational
thermal conductivity based on
the approximation (\ref{Phi}) with $\tau_c$ independent of $\varepsilon_c$.
However, the actual energy dependence of the function $\Phi_c$
in the exact solution is more complicated. As already mentioned
above,
Gnedin and Yakovlev \cite{gy95} introduced a factor
$C$ which corrects the variational thermal conductivity,
$\kappa_\mathrm{exact}=C\kappa_\mathrm{var}$.
For the electromagnetic interaction wholly (and erroneously)
attributed to the exchange
of longitudinal plasmons, in the weak-screening
approximation they obtained $C\approx 1.2-1.3$
and suggested to use $C=1.2$. Their calculation
was based on the exact transport theory in Fermi liquids
developed by Sykes
and Brooker \cite{sybr70} for one-component systems and extended
for multi-component systems by Flowers and Itoh \cite{fi79} and
Anderson {\it et al}.\ \cite{and87}.
Notice that if Gnedin and Yakovlev correctly
calculated the thermal conductivity solely determined
by the exchange of longitudinal plasmons they would
obtain the parameter $\lambda$ from their Eq.\ (62)
equal to 1 and $C=1.2$.
The exact theory \cite{sybr70}
cannot be directly applied to our case
because the Landau damping of transverse plasmons
makes the matrix element $\omega$-dependent, that is not
characteristic for ordinary Fermi liquids.

In this section we estimate $C$ taking into account only the exchange of
transverse plasmons in the
weak screening approximation for one type of heat
carriers (recall that the electron and muon heat transports
become decoupled). We use the straightforward
generalization of the exact theory
\cite{sybr70,fi79,and87,Baym91}. Instead of
Eq.\ (\ref{Phi}) we write
\begin{equation}
     \Phi_c= -\tau_\mathrm{eff}\Psi(x)\, \bm{v}_c \cdot
     \bm{\nabla} T ,
\label{Phi_exact}
\end{equation}
where $\tau_\mathrm{eff}$ is an effective relaxation time,
while $\Psi(x)$ is
an unknown function of $x=(\varepsilon-\mu_c)/T$.
Following Ref.\ \cite{Baym91} we substitute
(\ref{Phi_exact}) into the linearized kinetic equation and obtain the
equation for $\Psi(x)$,
\begin{eqnarray}
  &-&\frac{x}{1+\exp(x)}
\nonumber \\  
  &&=\int\limits_{-\infty}^\infty {\rm d} x_1\;
  \frac{ \sgn(x_1-x)}{1+\exp(x_1)}\;
  \frac{\Psi(x_1)-\Psi(x)}{1-\exp(x-x_1)},
\label{Psi}
\end{eqnarray}
with $\tau_\mathrm{eff}=3/(\alpha v_{Fc}T)$, and
the thermal conductivity
calculated as
\begin{equation}
  \kappa_{\mathrm{exact}}=\frac{n_c\tau_\mathrm{eff}}{m_c^*}
  \int\limits_{-\infty}^{\infty} {\rm d} x\
  x\Psi(x) f_c (1-f_c).
\label{kappa-exact}
\end{equation}

We have solved Eq.\ (\ref{Psi}) numerically,
calculated $\kappa_{\mathrm{exact}}$, and compared it
with the variational
result (that corresponds to
$\Psi_\mathrm{var}(x)=\pi^2 x/[18\zeta(3)]$).
The difference between the exact and variational results
appears to be
negligible, $C=\kappa_\mathrm{exact}/\kappa_\mathrm{var}\approx 1$.
In a more general case the electron (muon)
scattering is determined by the exchange of both transverse and
longitudinal plasmons
and we can expect that $1 \leq C \leq 1.2$.

\section{Effects of proton superconductivity}
\label{sect-superfluid}

In the previous sections we have considered
nonsuperconducting protons. However, it is well known
that the protons in neutron star cores can
be in a superfluid (superconducting) state (e.g., Lombardo
and Schulze \cite{ls01}). Let us include the effects
of proton superconductivity. 

Microscopically, the onset of superconductivity after the
temperature $T$ falls below a (density dependent)
critical temperature $T_{cp}$ manifests
itself in the appearance of a gap $\Delta=\Delta(T)$ in the
proton energy spectrum in the vicinity of the
proton Fermi level. Instead of protons,
it is now necessary to introduce Bogoliubov quasiparticles
with energies (with respect to the Fermi level $\mu_p$),
\begin{equation}
    \epsilon-\mu_p = \mathrm{sign}(p-p_{Fp})
    \sqrt{ \Delta^2 + v_{Fp}^2 (p-p_{Fp})^2}.
\label{BCS}
\end{equation}
It is generally believed, that
proton pairing in neutron star cores occurs in the $^1S_0$
channel. In this case, the  gap is
isotropic (independent of orientation of proton momentum with
respect to a quantization axis). For the Bardeen-Cooper-Schrieffer (BCS)
pairing model
the temperature dependence of
$\Delta$ at
$0 \leq T \leq T_{cp}$
can be fitted as \cite{ly94}
\begin{equation}
\label{E:delta0}
     y
     \equiv
     \frac{\Delta}{k_B T}=
     \sqrt{1-t}
     \left(1.456-\frac{0.157}{\sqrt{t}}+\frac{1.764}{t}\right),
\end{equation}
where $t=T/T_{cp}$.

\subsection{Dielectric function}
\label{dielectricfun}

\begin{figure}
\begin{center}
\leavevmode
\epsfxsize=80mm
\epsfbox[40 175 565 680]{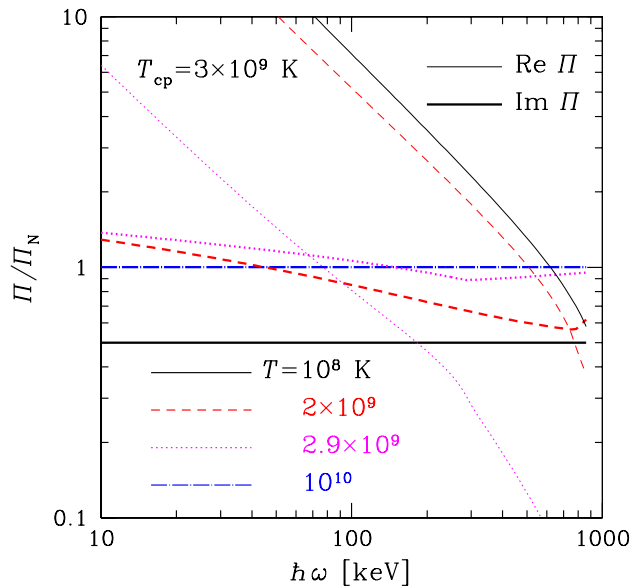}
\end{center}
\caption{(Color online) Real and imaginary parts 
    (thin and thick lines, respectively)
         of the proton transverse polarization function $\Pi_t^{(p)}$
     versus $\hbar \omega$ at small $q$ in superconducting matter
      for the proton
     critical temperature $T_{cp}=3\times 10^9$~K and 
     four
     values of the temperature, $T=10^8$ K (solid lines),
     $2\times 10^9$ K (dashed lines),
         $2.9\times 10^9$ K (dotted lines), and $10^{10}$ K
     (dot-dashed lines). The plotted functions are
     divided by the imaginary part of 
     $\Pi_{t\mathrm{N}}^{(p)}$ in normal matter.}
\label{fig:polariz}
\end{figure}

Proton superconductivity affects electrodynamical properties of
dense matter. The longitudinal polarization function
$\Pi_l$ undergoes no significant changes (see, e.g., Ref.\ \cite{alf06};
a strong effect of superconductivity
on longitudinal plasma screening described by the function $Z$ in
Eq.\ (47) of Ref.\ \cite{gy95} is in error;
one should set $Z=1$ in equations of Ref.\ \cite{gy95}).
In contrast, the proton component
$\Pi_t^{(p)}$ of the total transverse function
$\Pi_t=\Pi_t^{(e)}+\Pi_t^{(\mu)}+\Pi_t^{(p)}$ is considerably modified.
It should be stressed that we discuss the electromagnetic
polarization functions to be distinguished from
polarization functions associated with strong interactions
(e.g., Ref.\ \cite{kr04}).

For our purpose, we employ the model of
a Fermi gas of protons with BCS 
superconductivity
in
the so called Pippard limit 
appropriate
to neutron star cores
($\hbar \omega \ll p_{Fp}v_{Fp}$,
$\hbar q \ll p_{Fp}$, but $\xi q \gg 1$, with
$\xi \sim \hbar v_{Fp}/k_B T_{cp}$
being the coherence length). 
The polarization functions appropriate to this case are well
known. They
were calculated in the classical paper by Mattis and Bardeen \cite{Mattis58}.
In the limit of small $\omega$ and $q$, we are interested in, these
results are gauge invariant \cite{alf06}.
The effects of strong (nuclear) interactions in the proton Fermi liquid
(particularly, dragging of neutrons by protons) 
on the electromagnetic polarization functions 
are complicated and almost not explored; they are
beyond the scope of the present paper. According to 
M.~E.\ Gusakov (private communication) these effects should not
greatly modify our results. Similar effects of strong
interactions on the polarization functions of superfluid neutrons
\cite{kr04} lead to renormalization of vertices and take into
account specific collective superfluid excitations.
These effects can be pronounced for those $q$ and $\omega$
which are typical of such excitations ($\omega \sim q v_F$), 
and the effects are small
outside such $q-\omega$ domains. We expect that in our case
($\omega \ll q v_F$) these effects are small and can be neglected.

Therefore, using the
classical polarization functions \cite{Mattis58} seems to be
a good starting approximation. It gives  
%
\begin{equation}
     \Pi_t^{(p)}=\frac{q_t^2}{4} \, \frac{\Delta}{q} \, Q,
\end{equation}
where $Q$ is a complex function of
the normalized frequency $\tilde{\omega}=\hbar \omega/\Delta$
and the normalized gap $y$ (or, alternatively, of
$\hbar \omega/k_B T$ and $y$).
The real and imaginary parts of $Q$ are given by
\begin{eqnarray}
\Re
    Q&=&\pi\int\limits^{\tilde{\omega}+1}_{\max[\tilde{\omega}-1,1]}
    \tanh\left(\frac{y}{2}\,E\right)
\nonumber \\    
   && \times \frac{[E(E-\tilde{\omega})+1]
    \, {\rm d} E}{\sqrt{(E^2-1)(1-(E-\tilde{\omega})^2)}},
\\
    \Im Q &=& \pi\int\limits_1^\infty \left\{\tanh\left(\frac{y}{2}\,
    (E+\tilde{\omega})\right)
    -\tanh\left(\frac{y}{2}\, E\right)\right\}
\nonumber \\    
  && \times  \frac{[E(E+\tilde{\omega})+1]
    \, {\rm d}E}{\sqrt{(E^2-1)((E+\tilde{\omega})^2-1)}}
\nonumber\\
    &+& \pi
    \Theta(\tilde{\omega}-2)\int\limits_1^{\tilde{\omega}-1}
    \tanh\left(\frac{y}{2}\,E\right)
\nonumber \\    
    && \times \frac{[E(\tilde{\omega}-E)-1]
    \, {\rm d}E}{\sqrt{(E^2-1)((\tilde{\omega}-E)^2-1)}},
\end{eqnarray}
where $\Theta(x)=0$ at $x\leq 0$ and $\Theta(x)=1$ otherwise.
Here and below we again return to units $c=\hbar=k_B=1$
unless the contrary is indicated.

In the nonsuperconding limit, $y\to 0$, only the imaginary part of
$Q$ survives, giving $Q=i \pi \tilde{\omega}$ and
reproducing the asymptote (\ref{Pit}) determined by the
Landau damping (dissipative plasma screening). In the
opposite limit of strong superconductivity, $y \to \infty$, the imaginary
part of $Q$ is small (because the gap in the
proton energy spectrum becomes much larger
than characteristic values of $\hbar \omega \sim k_BT$).
Then only nondissipative plasma screening is available 
(with $Q=\pi^2$).
These results are illustrated in Fig.\ \ref{fig:polariz}
which shows the frequency dependence of the
real and imaginary parts of 
$\Pi_t^{(p)}$
in superconducting matter. 
The presented functions are normalized by the imaginary
(Landau damping) part of $\Pi_t^{(p)}$ in normal
matter. The critical temperature of proton superconductivity
is taken to be $T_{cp}=3 \times 10^9$~K
($k_BT_{cp}=259$ keV). The frequency
dependence is presented for four values of the temperature $T$.
The highest temperature $T=10^{10}$~K 
($k_BT=862$ keV) refers to nonsuperconducting
matter ($y=0$, $\Delta=0$). The proton polarization is provided by the Landau
damping. The displayed normalized polarization function
has the imaginary Landau-damping part (the thick dot-dashed line which is
the reference line of the figure); the real part is essentially zero.
The dotted, dash-dot, and solid lines refer to progressively
lower $T=2.9 \times 10^9$, $2 \times 10^9$, and
$10^8$~K ($k_BT=250$, 172, and 8.62 keV)
in superconducting matter (with
$y=0.570$, 2.26, and 52.6; and
$\Delta=142$, 389, and 453 keV, respectively). One can observe the growth
of the real part of the normalized polarization function,
which becomes larger than the imaginary part, leading
to a nondissipative transverse plasma screening for $T \ll T_{cp}$.

The total transverse polarization function of the
plasma containing electrons, muons and superconducting
protons has the form
\begin{eqnarray}
  \Pi_{t}&=&\frac{\pi\omega}{4q}\left\{
  q_{tp}^2\,\frac{\Delta}{\pi\omega}\,\Re
  Q \right.
\nonumber \\  
  &+&i \left. \left(q_{te}^2+q_{t\mu}^2 +
  q_{tp}^2\,\frac{\Delta}{\pi\omega}\,\Im Q
\right)\right\},
\end{eqnarray}
where $q_{ti}^2= 4 \alpha p_{Fi}^2/\pi$.
In the case of strong superconductivity, the main contribution to
transverse screening comes from superconducting protons.
When $T$ decreases from $T_{cp}$ to zero,
the transverse plasma screening turns from dissipative
to nondissipative one.

Thus, proton superconductivity modifies
the transverse polarization function and the screening functions
(\ref{fper}) and (\ref{fcomb}).
One can show that in the weak screening approximation
the angular integrals in superconducting matter can be presented as
\begin{eqnarray}
  I_{\Omega 1}^{\perp S} &=&
  I_{\Omega 1}^{\perp} F^\perp(\omega/T,y,r),
\\
  I_{\Omega 2}^{\perp \parallel S} &=&
  I_{\Omega 2}^{\perp \parallel} F^{\perp\parallel}(\omega/T,y,r),
\end{eqnarray}
where $I_{\Omega 1}^{\perp}$ and  $I_{\Omega 2}^{\perp\parallel}$ are
the angular integrals in the nonsuperconducting matter, while
\begin{eqnarray}
   F^\perp
   &=&\frac{\pi|\omega|(r+1)}{|\pi\omega
   r+\Delta\Im Q|}\,
\nonumber \\   
   && \times \left(1-\frac{2}{\pi}\arctan
   \frac{\Delta\Re Q}{|\pi\omega r+\Delta\Im
   Q|} \right),
\\
   F^{\parallel\perp}&=&
   \frac{\left[\pi|\omega|(r+1)\right]^{1/3}\left[\left(\pi\omega
   r+\Delta\Im Q\right)^2+ \left(\Delta\Re
   Q\right)^2\right]^{1/3}}{|\pi\omega r+\Delta\Im
   Q|}
\nonumber
\\
   &\times&\frac{2}{\sqrt{3}}\sin\left(\frac{2}{3}\arctan\frac{|\pi\omega
   r+\Delta\Im Q|}{\Delta\Re Q}
   \right)
\nonumber
\\
   &+&\frac{\left[\pi|\omega|(r+1)\right]^{1/3}\Delta\Re
   Q}{|\pi\omega r+\Delta\Im
   Q|\left[\left(\pi\omega r+\Delta\Im
   Q\right)^2+ \left(\Delta\Re
   Q\right)^2\right]^{1/6}}
\nonumber
\\
   &\times&\frac{2}{\sqrt{3}}\sin\left(\frac{1}{3}\arctan\frac{|\pi\omega
   r+\Delta\Im Q|}{\Delta\Re Q}\right)
\end{eqnarray}
are the factors which take into account the effects of superconductivity.
The parameter $r$ is defined as
\begin{equation}
     r=\frac{q_{te}^2+q_{t\mu}^2}{q_{tp}^2}
     =\frac{p_{Fe}^2+p_{F\mu}^2}{p_{Fp}^2}.
\end{equation}
It is a slowly variable function of the density in a neutron star
core, constrained by the condition of plasma neutrality,
$n_e+n_\mu=n_p$. We have $r=1$ in the absence of muons, and $r>1$
if the muons are present. The maximum value $r_{\rm max}= 2^{1/3} \approx
1.26$ is reached in the high-density limit in which the muons
become ultrarelativistic.

In the limiting case of strong superconductivity ($y \gg 1$)
\begin{eqnarray}
    F^\perp&=&\frac{2|\omega|(r+1)}{\pi^2\Delta},
\\
    F^{\parallel\perp}&=&\frac{2}{\sqrt{3}}
    \left(\frac{|\omega|(r+1)}{\pi\Delta}\right)^{1/3}.
\end{eqnarray}

The collision frequencies involving nonsuperconducting particles
($ee$, $\mu\mu$, and $e\mu$) can be
calculated directly from Eqs.\ (\ref{Jci1}) and (\ref{Jci2}), taking
into account the factors $F^\perp$ and $F^{\parallel\perp}$. It is
convenient to present these collision frequencies in the form
\begin{eqnarray}
   \nu^{\perp S}_{ci}&=&\nu^{\perp}_{ci}\, R_\ell^\perp(y,r),
\\
   \nu^{' S}_{ci}&=&\nu'_{ci}\, R_\ell^{\parallel\perp}(y,r),
\end{eqnarray}
where $\nu^{\perp}_{ci}$ and $\nu'_{ci}$ correspond to
nonsuperconducting matter, while
\begin{eqnarray}
   R_\ell^\perp(y,r)&=& \frac{1}{6\zeta(3)}\int_0^\infty
   {{\omega^3} \exp(-\omega/T) \,  F^\perp\,{\rm d}\omega \over
   \left[1- \exp(-\omega/T)\right]^2},
\nonumber \\
   R_\ell^{\parallel\perp}(y,r)&=&
   \frac{1}{\Gamma(14/3)\zeta(11/3)}
\nonumber \\   
  && \times \int_0^\infty
   {{\omega^{11/3}} \exp(-\omega/T) \,
   F^{\parallel \perp}\,{\rm d}\omega \over
   \left[1-\exp (-\omega/T)\right]^2}
\label{R-perp-l}
\end{eqnarray}
describe the effect of proton superconductivity.
In the  case of strong superconductivity ($y \gg 1$) we
obtain the asymptotes
\begin{eqnarray}
    R_\ell^\perp(y,r)&=&\frac{4\pi^2(r+1)}{45\zeta(3)y},
\nonumber \\
    R_\ell^{\parallel\perp}(y,r)&=&\frac{8\pi^4}
    {\Gamma(14/3)\zeta(11/3)15\sqrt{3}}
    \left(\frac{r+1}{\pi y}\right)^{1/3}.
\label{R-perp-l-asy}
\end{eqnarray}

We have calculated $R_\ell^{\parallel\perp}(y,r)$ on a dense
grid of $y$ and $r$ and fitted the results
by the function
\begin{equation}
   R_\ell^{\parallel\perp}(y,r)=\frac{(r+1)^{1/3}}
   {\left[(r+1)^2-0.757 y +(0.50651y)^2\right]^{1/6}}.
\label{Jfit}
\end{equation}
The fit reproduces numerical data, the above asymptote at $y \to \infty$,
and the evident condition $R_\ell^{\parallel\perp}(0,r)=1$.
The maximum relative fit error 
$\approx 8$\%
occurs at $y=2.2667$ and
$r=1.3$. We do not present any separate fit formula
for $R_\ell^\perp(y,r)$ but obtain a practical expression for
$\nu^{\perp S}_c$ in the next section.

\subsection{Superconducting scatterers}
\label{superfluscatter}

Aside of changing plasma screening, proton
superconductivity directly affects the partial collision frequencies
$\nu_{ep}$ and $\nu_{\mu p}$. Now these collision frequencies
describe interaction of electrons or muons
with proton Bogoliubov quasiparticles. Because the number
of such quasiparticles is not necessarily conserved,
the collision frequencies should include contributions
of three processes, (1) \textit{scattering}
($1,2 \to 1',2'$); (2) \textit{decay} ($1 \to 1',-2,2'$);
and (3) \textit{coalescence} ($1,2,-2' \to 1'$).
Here, $1$ and $1'$ refer to an electron or muon;
$2$, $2'$, $-2$, and $-2'$ refer to Bogoliubov
proton quasiparticles; $-2$ (or $-2'$) stands for a proton quasiparticle
with momentum and spin directions opposite to those
for $2$ (or $2'$). A linearized collision integral $I_{cp}$ becomes
\begin{equation}
\label{collint_sf}
   I_{cp}=I^\mathrm{sc}_{cp}+I^\mathrm{dec}_{cp}+I^\mathrm{coal}_{cp},
\end{equation}
where
\begin{eqnarray}
   I^\mathrm{sc}_{cp}&=&\int \mathrm{d}\Gamma\,
   f_1f_2(1-f_{1'})(1-f_{2'})
\nonumber \\   
   && \times W^\mathrm{sc}_{cp}(1,2|1',2')
   \left(\Phi_{1'}-\Phi_1\right),
\label{collint_sc}   \\
   I^\mathrm{dec}_{cp}&=&\frac{1}{2}\int \mathrm{d}\Gamma\,
   f_1(1-f_{-2})(1-f_{1'})(1-f_{2'})
\nonumber \\   
   &&\times   
   W^\mathrm{dec}_{cp}(1|1',-2,2')
   \left(\Phi_{1'}-\Phi_1\right),
\label{collint_dec}\\
   I^\mathrm{coal}_{cp}&=&\frac{1}{2}\int \mathrm{d}\Gamma\,   
   f_1f_2f_{-2'}(1-f_{1'})
\nonumber \\
   && \times    
   W^\mathrm{coal}_{cp}(1,2,-2'|1')
   \left(\Phi_{1'}-\Phi_1\right),
\label{collint_coal}
\end{eqnarray}
correspond to scattering (sc), decay (dec), and coalescence (coal),
respectively;
$\mathrm{d}\Gamma=(2\pi\hbar)^{-9}
\mathrm{d}^3p_2\,\mathrm{d}^3p_{1'}\,\mathrm{d}^3p_{2'}$.
The factor $1 \over 2$ excludes double counting of
the same collision events.

Let us focus on the differential transition probabilities
$W^\mathrm{sc}_{cp}(1,2|1',2')$,
$W^\mathrm{dec}_{cp}(1|1',-2,2')$, and
$W^\mathrm{coal}_{cp}(1,2,-2'|1')$.
In the case of scattering, according
to Fermi Golden Rule,
\begin{equation}
   W_{cp}^\mathrm{sc}= 2\pi
   \delta(\varepsilon_1+\varepsilon_2
   -\varepsilon_{1'}-\varepsilon_{2'})
   \sum_\mathrm{spins}
   \left|\langle 1',2' |\hat{V} |1,2\rangle\right|^2,
\label{Wsc}
\end{equation}
where
\begin{eqnarray}
    \hat{V}&=&\alpha \int\int
    \mathrm{d}\bm{r}_1\, \mathrm{d}\bm{r}_2\,
    \left[
    \hat{\rho}_c(\bm{r}_1)\hat{\rho}_p(\bm{r}_2)
    D_\parallel(\bm{r}_1-\bm{r}_2) \right.
\nonumber \\    
    && - 
    (\hat{\bm{J}}_c(\bm{r}_1)\cdot\hat{\bm{J}}_p(\bm{r}_2))
    \left. D_\perp(\bm{r}_1-\bm{r}_2)\right]
\end{eqnarray}
is the interaction operator,
$D_\parallel$ and $D_\perp$ being corresponding propagators.
In the first approximation, protons in neutron star cores can be treated as
nonrelativistic particles.
Then the
second-quantized operators of the proton density and
proton current are
\begin{eqnarray}
    \hat{\rho}_p&=&\hat{\psi}^\dag\hat{\psi},
\label{dens}\\
    \hat{\bm{J}}_p&=&\frac{i}{2m^*_p}
    \left[(\bm{\nabla} \hat{\psi}^\dag)\hat{\psi}
    -\hat{\psi}^\dag\bm{\nabla}\hat{\psi}\right].
\label{curr}
\end{eqnarray}

The psi-operator of proton field is
\begin{equation}
  \hat{\psi}=\sum_{\bm{p}\sigma}
  \chi_{\sigma}\exp(i\bm{p}\cdot\bm{r})\,
  \hat{a}_{\bm{p}\sigma},
\label{psip}
\end{equation}
where
$\hat{a}_{\bm{p}\sigma}$ is a proton
(not 
a
Bogoliubov quasi-particle) annihilation operator and
$\chi_\sigma$ is a 
unit
basic spinor. Bogoliubov
quasiparticles are introduced through
quasiparticle annihilation operators $\hat{b}_{\bm{p}\sigma}$
basing on the Bogoliubov transformation
\begin{eqnarray}
   \hat{b}_{\bm{p}\sigma}&=&\mathrm{u}_{p}
   \hat{a}_{\bm{p}\sigma}-\sgn(\sigma)\,
   \mathrm{v}_{p}\hat{a}^\dag_{-\bm{p},-\sigma},
\nonumber   \\
   \hat{a}_{\bm{p}\sigma}&=&\mathrm{u}_{p} \hat{b}_{\bm{p}\sigma}+
   \sgn(\sigma)\,\mathrm{v}_{p}\hat{b}^\dag_{-\bm{p},-\sigma},
\label{bogoliubov}
\end{eqnarray}
where $\mathrm{u}_p$ and $\mathrm{v}_p$ are coherence factors
which can be chosen in different ways. We will
use the following set of these factors
\begin{eqnarray}
  \mathrm{u}_p&=&\frac{1}{\sqrt{2}}\,\sqrt{1+\frac{x}{z}},
\nonumber  \\
  \mathrm{v}_p&=&\frac{\sgn(x)}{\sqrt{2}}\,\sqrt{1-\frac{x}{z}},
\label{uv}
\end{eqnarray}
where $x=v_{Fp}(p-p_{Fp})/T$ and $z=(\varepsilon-\mu_p)/T$.
In the nonsuperconducting limit we have  $\mathrm{u}_p(y=0)=1$ and
$\mathrm{v}_p(y=0)=0$.
With this choice
the dimensionless energy spectrum of quasiparticles is
\begin{equation}
\label{quasispectrum}
    z={\varepsilon-\mu_p \over T}=\sgn(x) \,\sqrt{x^2+y^2}.
\end{equation}
Obviously, the introduction of proton
quasiparticles affects only the proton part of the 
matrix element that describes electromagnetic interaction.
For the longitudinal part we have
\begin{equation}
   \langle 2'|\hat{\rho}_p|2\rangle =
   (\mathrm{u}_{2'}\mathrm{u}_2-\mathrm{v}_{2'}\mathrm{v}_2)\,
   \delta_{\sigma_2\sigma_{2'}}
   \exp(i (\bm{p}_{2'}-\bm{p}_2)\cdot\bm{r}).
\label{dens_matel}
\end{equation}
%
It is clear that the longitudinal component of the
matrix element $M^\mathrm{sc}_\parallel$ differs from
the same component 
$M^\mathrm{norm}_\parallel$
in a nonsuperconducting (normal) case by
\begin{eqnarray}
   M^\mathrm{sc}_\parallel &=&
   \left( \mathrm{u}_{2'}\mathrm{u}_{2}-\mathrm{v}_{2'}\mathrm{v}_{2}\right)
   M^\mathrm{norm}_\parallel.
\label{mat-elem-parall}
\end{eqnarray}
In the same way for the matrix element of the proton current
$\langle 2'|\hat{\bm{J}}_p|2\rangle$, that describes the transverse
interaction, we obtain
\begin{equation}
    M^\mathrm{sc}_\perp =
    \left( \mathrm{u}_{2'}\mathrm{u}_{2}+
    \mathrm{v}_{2'}\mathrm{v}_{2}\right)M^\mathrm{norm}_\perp.
\label{mat-elem-transers}
\end{equation}

The differential transition probability
for the decay channel has the form
\begin{equation}
    W_{cp}^\mathrm{dec}= 2\pi
    \delta(\varepsilon_1-\varepsilon_2-\varepsilon_{1'}-\varepsilon_{2'})
    \sum_\mathrm{spins}\left|\langle 1',-2,2'|\hat{V} |1\rangle\right|^2.
\label{Wdecay}
\end{equation}
Notice that the energy conserving delta function differs from that
in the scattering process. However, momentum and spin
selection restrictions are the same for scattering,
decay and coalescence processes because of the adopted choice of
initial and final particle momenta and spin quantum numbers.
We obtain
\begin{eqnarray}
    M^\mathrm{dec}_{\parallel} &=&
    \left( \mathrm{u}_{2'}\mathrm{v}_{2}+\mathrm{u}_{2}\mathrm{v}_{2'}\right)
    \sgn(\sigma_{2})M^\mathrm{norm}_{\parallel},
\nonumber    \\
    M^\mathrm{dec}_{\perp} &=&
    \left( \mathrm{u}_{2'}\mathrm{v}_{2}-\mathrm{u}_{2}\mathrm{v}_{2'}\right)
    \sgn(\sigma_{2})M^\mathrm{norm}_{\perp}.
\label{decay-mat-elem}
\end{eqnarray}

Similarly, for the coalescence channel
\begin{eqnarray}
    W_{cp}^\mathrm{coal}&=& 2\pi
   \delta(\varepsilon_1+\varepsilon_2-\varepsilon_{1'}+\varepsilon_{2'})
\nonumber \\   
   && \times
   \sum_\mathrm{spins}\left|\langle1'|\hat{V} |1,2,-2'\rangle\right|^2,
\label{Wcoal}   \\
   M^\mathrm{coal}_{\parallel}
    &=&\left( \mathrm{u}_{2'}\mathrm{v}_{2}
    +\mathrm{u}_{2}\mathrm{v}_{2'} \right)\sgn(\sigma_{2})
    M^\mathrm{norm}_{\parallel},
\nonumber \\
   M^\mathrm{coal}_{\perp} &=&
   \left( \mathrm{u}_{2}\mathrm{v}_{2'}-
   \mathrm{u}_{2'}\mathrm{v}_{2}\right)\sgn(\sigma_{2})
   M^\mathrm{norm}_{\perp}.
\label{norm-mat-elem}
\end{eqnarray}
Our procedure is standard and tacitly
ignores renormalization of proton vertices in superconducting matter.
It is justified because we
are interested is small energy and momentum transfers
$\omega$ and $q$ which are far from the transfers typical of
collective superfluid excitations (see Refs.\ \cite{kr04,alf06} 
and Sec.\ \ref{dielectricfun}).

While integrating over quasiproton momenta in collision integrals
we set $\mathrm{d}^3p=m_p^* T p_{Fp}\, \mathrm{d}x\, \mathrm{d}\Omega$.
It is instructive to replace $x_2\to -x_2$ for the decay case, and
$x_{2'} \to -x_{2'}$ for coalescence. Such replacements change
sign of quasiproton energy $(\varepsilon-\mu_p)$ and coherence factor
$\mathrm{v}_p$ because of our choice of these factors. Then all
energy-dependent terms, except for coherence factors, become the
same in all three collision integrals. Hence we can describe the
interaction of electrons (or muons) and quasiprotons by a unified
collision integral (\ref{collint_sc}), where
$W_{cp}^\mathrm{sc}(1,2|1',2')$ is substituted by
$W_{ep}^{S}(1,2|1',2')$,
\begin{eqnarray}
   &&  W_{cp}^{S}(1,2|1',2')=4(2\pi)^4
     \delta(\varepsilon_1+\varepsilon_2-\varepsilon'_1 - \varepsilon'_2)
\nonumber \\     
    && \qquad \times \delta( \bm{p}_1+\bm{p}_2-\bm{p}'_1-\bm{p}'_2)\,|M^{S}_{cp}|^2,
\label{Wunified} \\
    && |M^{S}_{cp}|^2=|M^\mathrm{sc}_{cp}|^2
\nonumber \\    
 && \qquad     +\frac{1}{2}\,|M^\mathrm{dec}_{cp}(-x_2)|^2
      +\frac{1}{2}\,|M^\mathrm{coal}_{cp}(-x_2')|^2.
\label{Mat-elem-unified}
\end{eqnarray}

Putting the squared matrix elements together, we have
\begin{eqnarray}
     \varphi^S_\parallel&=&
     (1-4\mathrm{u}_2\mathrm{u}_{2'}
     \mathrm{v}_2\mathrm{v}_{2'})\,\varphi_\parallel,
\nonumber \\
     \varphi^S_\perp&=&
     (1+4\mathrm{u}_{2}\mathrm{u}_{2'}
     \mathrm{v}_2\mathrm{v}_{2'})\,\varphi_\perp,
\nonumber  \\
     \varphi^S_{\parallel\perp}&=&
     (1-\mathrm{v}_2^2-\mathrm{v}_{2'}^2)\,\varphi_{\parallel\perp},
\label{phis}
\end{eqnarray}
instead of $\varphi_\parallel$, $\varphi_\perp$, and
$\varphi_{\parallel\perp}$
given by Eqs.\ (\ref{fpar})--(\ref{fcomb}) in the nonsuperconducting
case.

Therefore, the expressions for $\nu_{ep}$ or $\nu_{\mu p}$
in the presence of proton superconductivity can be presented
in the integral form similar to that in the nonsuperconducting case,
but with two differences. First, coherence factors have to
be introduced in the squared matrix element in
accordance with Eq.\ (\ref{phis}). Second, proper
quasiproton energies (\ref{quasispectrum}), containing energy gaps,
should be used in Fermi-Dirac distribution functions of
quasiprotons. Thus in order to calculate
$\nu_{ep}$ and $\nu_{\mu p}$ we should reconsider not only the
angular integration due to changes in polarization functions but
the energy integration as well. The collision frequencies in
question contain the transverse and longitudinal components
(\ref{nuci}). A careful examination of the derivation of the
energy integral in (\ref{Jci1}) shows that for the collisions via
the exchange of transverse plasmons one has
\begin{eqnarray}
   \nu_{cp}^{\perp S}&=&\nu_{cp}^{\perp}\,R^\perp_p(y,r), 
\label{nu-cp-perp}\\
    R^\perp_p(y,r)&=&\frac{1}{6\zeta(3)}
    \int\limits_0^\infty\int\limits_0^\infty
    \frac{{\rm d}x_2\, {\rm d}x_2'}{1+\exp(z_2)} 
\nonumber \\
 &&   \times \left\{ \frac{(z_2'-z_2)|z_2'-z_2|
    F^\perp(|z_2'-z_2|,y,r)}{[1+\exp(-z_2')][\exp(z_2'-z_2)-1]}
    \right.
\nonumber \\    
 &&  \times (1+
    4\mathrm{u}_2\mathrm{u}_2'\mathrm{v}_2\mathrm{v}_2') 
\nonumber\\
&&  -
     \left.\frac{(z_2'+z_2)|z_2'+z_2|
     F^\perp(|z_2'+z_2|,y,r)}{[1+\exp(z_2')][\exp(-z_2'-z_2)-1]}
     \right.
\nonumber \\     
 && \times 
 \left. {{}\over{}}  
 \left(1-4\mathrm{u}_2\mathrm{u}_2'\mathrm{v}_2\mathrm{v}_2'\right)
     \right\},
\nonumber 
\end{eqnarray}
where $\nu_{cp}^\perp$ corresponds to normal protons,
$R_p^\perp(y,r)$ describes the effects of superconductivity,
$z_2= (\epsilon_2-\mu_p)/k_BT=
\sqrt{x_2^2+y^2}$ and 
$z'_2=(\epsilon'_2-\mu_p)/k_BT=\sqrt{x^{\prime 2}_2+y^2}$.

In the limit of strong superconductivity ($y\gg 1$) we have
$R_p^{\perp}=A_\perp(r+1) \exp(-y)$, where
\begin{eqnarray}
   A_\perp&=&\frac{4}{3\pi^2\zeta(3)}\int_0^\infty 
   {\rm d}\eta_1 \int_0^\infty {\rm d}\eta_2
\nonumber \\
   && \times \frac{(\eta_1^2-\eta_2^2)^3}
                   {\exp(\eta_1^2)-\exp(\eta_2^2)}\approx 0.3446.
\end{eqnarray}
This strong superconductivity exponentially reduces
the $ep$ and $\mu p$ collision rates. Nevertheless, the protons
give the main contribution to the
plasma polarization associated with the
exchange of transverse plasmons. In this way they
remain vitally important for
the plasma screening in collisions involving leptons alone.

The total transverse collision frequency of heat carriers $c$ is
\begin{equation}
    \nu_c^{\perp S}=\sum_i \nu_{ci}^{\perp S} = \nu_c^{\perp}
    R_\mathrm{tot}^\perp(y,r),
\label{nu-perp-S-tot}
\end{equation}
where $\nu_c^\perp$ is the total collision frequency in nonsuperconducting
matter and
\begin{equation}
\label{Rperp}
  R_\mathrm{tot}^\perp(y,r)=\frac{1}{r+1}\,
  \left[rR_\ell^\perp(y,r)+R^\perp_p(y,r)\right]
\end{equation}
describes the overall superconducting suppression
of the total collision frequency via the exchange of transverse plasmons.
In the limit of strong superconductivity, we have
$R_\mathrm{tot}^\perp(y,r)=4\pi^2r/[45\zeta(3)y]$.

We have calculated $R_\mathrm{tot}^\perp(y,r)$ on a dense
grid of $y$ and $r$ and fitted the results by the formula
\begin{eqnarray}
   R_\mathrm{tot}^\perp(y,r)&=& p_1 \exp\left(-0.14 y^2\right)+
   \frac{1-p_1}{\sqrt{1+p_3y^2}},
\label{R-perp-fit}\\
   p_1&=&0.48-0.17r,
\nonumber \\
   p_3&=&\left[(1-p_1)\frac{45\zeta(3)}{4\pi^2r}\right]^2,
\nonumber
\end{eqnarray}
which reproduces also the asymptote of
$R_\mathrm{tot}^\perp(y,r)$ at $y \to \infty$ and
the obvious condition $R_\mathrm{tot}^\perp(0,r)=1$.
The maximum fit error $\approx 5$\%
takes place at $y=1.2$ and $r=1$.

For the longitudinal components of the
collision frequencies $\nu_{ep}^\parallel$
and $\nu_{\mu p}^\parallel$, the angular
integration in (\ref{IOmega1}) remains unchanged.
Then 
\begin{equation}
     \nu_{cp}^{\parallel S}=\nu_{cp}^{\parallel}\,R^\parallel_p(y),
\label{nu-parall-S}
\end{equation}
where $\nu_{cp}^\parallel$ is the nonsuperconducting value, while
\begin{eqnarray}
    R^\parallel_p(y)&=&\frac{15}{4\pi^4}
    \int\limits_0^\infty\int\limits_0^\infty
    \frac{\mathrm{d}x_2 \, \mathrm{d}x_2'}{1+\exp(z_2)}
\nonumber \\    
   && \times \left\{\frac{(z_2'-z_2)^3 
    (1-4\mathrm{u}_2\mathrm{u}_2'\mathrm{v}_2\mathrm{v}_2')}
    {[1+\exp(-z_2')][\exp(z_2'-z_2)-1]}
    \right.
\nonumber\\
&& -
     \left.\frac{(z_2'+z_2)^3 
     (1+4 \mathrm{u}_2\mathrm{u}_2'\mathrm{v}_2\mathrm{v}_2')}
     {[1+\exp(z_2')][\exp(-z_2'-z_2)-1]}
\right\}
\label{R-parall-S}
\end{eqnarray}
is the superfluid reduction factor.

In the limit of $y\gg 1$ we obtain
$R^\parallel_p(y)\to A_\parallel \exp(-y)$, where
\begin{eqnarray}
    A_\parallel&=&\frac{15}{2\pi^4}
    \int\limits_0^\infty \mathrm{d}\eta_1
    \int\limits_0^\infty \mathrm{d}\eta_2\;
\nonumber \\
    && \times
    \frac{(\eta_1^2-\eta_2^2)^3(\eta_1^2+\eta_2^2)}
    {\exp(\eta_1^2)-\exp(\eta_2^2)}\approx 1.00.
\label{A-parall}
\end{eqnarray}

Finally, in analogy with Eq.\ (\ref{R-perp-fit}),
we have numerically calculated $R^\parallel_p(y)$ and
fitted it by
\begin{eqnarray}
   R^\parallel_p(y) & = &
   \left\{1+\left(26.33y^2+0.376y^4\right){{}\over{}} \right.
\nonumber \\   
  & \times & \exp\left(-\sqrt{(3.675)^2+y^2}\right)
\nonumber\\
   & + & \left. 0.742\left[\exp\left((1.673)^2-\sqrt{(1.673)^2+y^2}\right)
   -1\right]
   \right\}
\nonumber\\
   & \times &  \exp\left((1.361)^2-\sqrt{(1.361)^2+y^2}\right)
\label{gnedin}
\end{eqnarray}
with the error $\lesssim 1\,\%$.

The reduction factor $R_p^\parallel(y)$ was introduced by Gnedin
and Yakovlev \cite{gy95} [see their Eq.\ (43)]. In contrast to our
Eq.\ (\ref{R-parall-S}), their reduction factor had two drawbacks.
First, it neglected coherence factors -- the terms containing 
$\mathrm{u}_2$ and
$\mathrm{v}_2$ in (\ref{R-parall-S}). Second, 
the authors of Ref.\ \cite{gy95} attributed the same
reduction $R_p^\parallel(y)$ to collisions via the exchange of
transverse and longitudinal plasmons [whereas in fact the
reduction factors $R_p^\perp(y,r)$ and $R_p^\parallel(y)$ are
different].
However, because superfluid reduction of $cp$ collisions
for $y \gg 1$ is exponentially strong, these drawbacks do not
change noticeably numerical values of the thermal conductivity
(while neglecting the exchange of transverse plasmons 
makes results really inaccurate).

Proton superconductivity significantly modifies the thermal
conductivity of electrons and muons. In particular, it violates
the temperature independence of the conductivity
(\ref{E:kappa_normal}) for nonsuperconducting matter.
The asymptote (\ref{Rperp}) indicates that
in a strongly superconducting matter ($T \ll T_{cp}$, $y \gg 1$)
the collision frequency $\nu_c^{\perp S}$ becomes much smaller than
in the nonsuperconducting case. Strong superconductivity
enforces the temperature dependence
$\nu_c^{\perp S}\propto T^2$
that is formally the same as for the longitudinal collision frequency
$\nu_c^{\parallel}$ in normal Fermi liquid.
Nevertheless, this thermal conductivity remains smaller
than the conductivity in the normal matter if it
included artificially the exchange of longitudinal plasmons alone.

As in normal matter, the major contribution
to the electron or muon thermal conductivity in the presence
of proton superconductivity
comes from the exchange of transverse plasmons.
Using the asymptotes for the reduction factors
in the limit of strong superconductivity ($y \gg 1$)
we obtain (in standard physical units)
\begin{equation}
   \kappa^{S}_c\approx\kappa^{S\perp}_c\approx
   \frac{5}{24}\,\frac{k_B c p_{Fc}^2}{\alpha \hbar^2}\,\frac{\Delta}{k_B T}\,
   \frac{p_{Fp}^2}{p_{Fe}^2+p_{F\mu}^2}.
\label{restoration}
\end{equation}
Therefore, strong proton superconductivity formally restores the
temperature dependence of the thermal conductivity $\kappa_{e\mu} \propto
1/T$ typical for a normal Fermi liquid (by enforcing
nondissipative plasma screening). Accordingly, we expect that in
this case our simplest variational thermal conductivity is
different from the exact conductivity by the same factor $C
\approx 1.2$ that is characteristic for Coulomb scattering via
the exchange of longitudinal plasmons \cite{gy95}. These arguments support our
expectations (Sec.\  \ref{sect-exact}) that
the overall uncertainties of our calculations of $\kappa_{e\mu}$ are
$\sim 20\%$. Notice that $\kappa_{e\mu}^S$ is proportional to
the superfluid gap $\Delta$.

\section{Results and Discussion}
\label{sect-results}

\subsection{Equations of state}
\label{sect-EOSs}

\begin{figure}
\begin{center}
\leavevmode
\epsfxsize=80mm
\epsfbox[40 175 565 680]{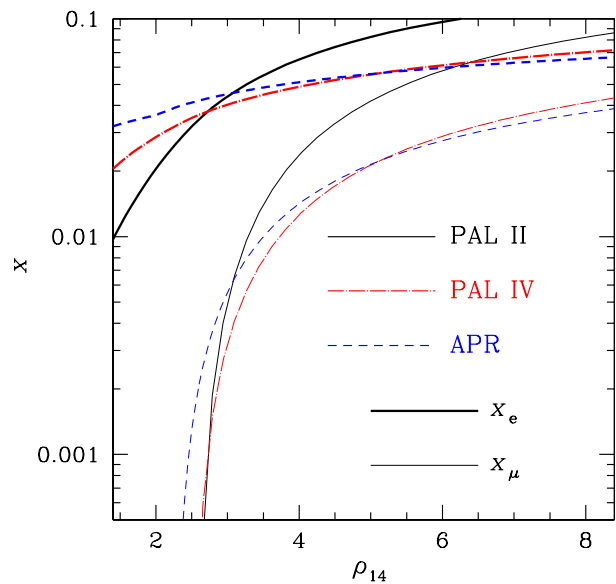}
\end{center}
\caption{(Color online)
  Fractions of electrons (thick lines) and
  muons (thin lines) versus density in a neutron star core for
  three selected EOSs --- APR (dashed lines), PAL II (solid lines), and
  PAL IV (dash-dot lines).}
\label{fig:frac}
\end{figure}

Let us illustrate our results by calculating the thermal
conductivity in a neutron star core composed of neutrons,
protons, electrons and muons. We employ three representative
equations of state (EOSs) which we
denote as APR, PAL II, and PAL IV. The fractions of
electrons $x_e=n_e/n_b$ and muons $x_\mu=n_\mu/n_b$
versus density of matter $\rho_{14}= \rho/
(10^{14}$ g~cm$^{-3})$ for these EOSs are plotted
in Fig.\ \ref{fig:frac} 
($n_b=n_n+n_p$ being the number density of baryons).
Let us recall that the lowest density in a neutron star core
is $\rho_\mathrm{min}\approx 1.5 \times 10^{14}$ g~cm$^{-3}$.

The APR EOS was constructed
by Akmal, Pandharipande and Ravenhall \cite{apr98}
(their model 
Argonne 
V18 + $\delta v$ + UIX$^*$). 
It
is sufficiently stiff and
gives the maximum gravitational
mass of stable neutron stars $M_\mathrm{max}=2.2\,M_\odot$.
The central density of the maximum-mass star is
$\rho_\mathrm{max} = 2.8 \times 10^{15}$ g~cm$^{-3}$.
No hyperons appear for this EOS at $\rho \leq \rho_\mathrm{max}$.
The threshold density
for the appearance of muons is
$\rho_{\mu}\approx2.28 \times 10^{14}$ g~cm$^{-3}$.

The PAL II is a convenient
phenomenological semianalytical model EOS proposed by Prakash,
Ainsworth and Lattimer \cite{pal88}.
The authors suggested several EOSs of such a type which
differ by a value of the compression modulus $K_0$
of saturated symmetric nuclear matter and by
the symmetry energy $S$ of dense matter as
a function of baryon number density $n_b$
[described by a function $F(u)$, where
$u=n_b/n_0$ and $n_0\approx 0.16$ fm$^{-3}$
is the baryon number density in saturated
symmetric nuclear matter].
The compression modulus determines the stiffness
of an EOS, while the symmetry energy
regulates the fraction of protons (and hence of electrons
and muons). The PAL II EOS
corresponds to $F(u)=2u^2/(u+1)$ which gives a rather high
fraction of protons at $u>1$ and the muon threshold density
$\rho_\mu=2.647 \times 10^{14}$ g~cm$^{-3}$ (Fig.\ \ref{fig:frac}).
Number densities
of various particles, and hence 
$\kappa_{e\mu}$, are insensitive to the value of $K_0$
for PAL EOSs \cite{pal88}.
Notice that by
taking $K_0$=120, 180, and 240 MeV (as suggested in Ref.\ \cite{pal88}),
we would obtain three modifications of the PAL II EOS with different
compressibility (from soft to stiff) which would give very
different neutron star models (different
mass-radius relations and maximum masses).

The PAL IV EOS belongs to the same PAL family of EOSs \cite{pal88} but
corresponds to the function $F(u)=u^{0.7}$ suggested
by Page and Applegate \cite{pa92}. It gives noticeably
lower fractions of protons, muons and electrons at
$u>1$ than those provided by the PAL II EOS (Fig.\ \ref{fig:frac}).
The muon creation threshold is $\rho_\mu = 2.51 \times 10^{14}$
g~cm$^{-3}$. Taking again different $K_0$ we
get different modifications of the PAL IV EOS,
from soft to stiff, giving different neutron star models.

Therefore, the selected EOSs
cover a large range of physical models of $npe\mu$
matter in neutron star cores.

\subsection{Thermal conductivity in nonsuperconducting cores}
\label{sect-nonsuperfluid}

\begin{figure}
\begin{center}
\leavevmode
\epsfxsize=80mm
\epsfbox[40 175 565 680]{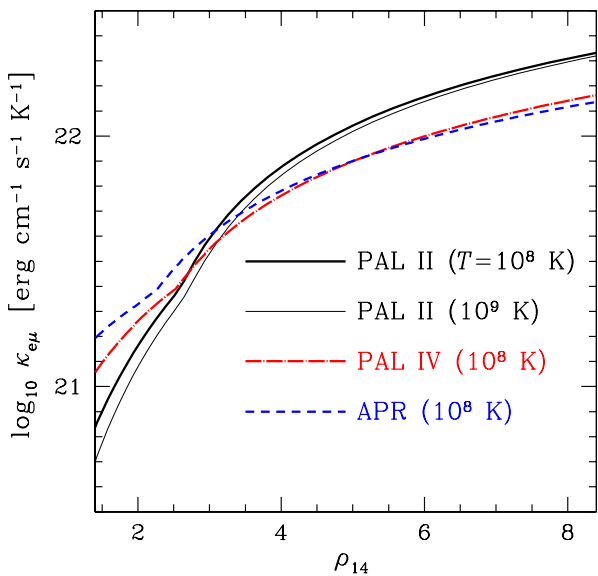}
\end{center}
\caption{(Color online) Electron-muon thermal conductivity
         $\kappa_{e\mu}$
     versus density in a nonsuperconducting neutron star
     core. Thick lines are for the APR (dashed line),
     PAL II (solid line) and PAL IV (dot-dashed line)
     EOSs at $T=10^8$~K. Thin solid line is for the PAL II
     EOS at $T=10^9$~K. }
\label{fig:kappaemu}
\end{figure}

Figure \ref{fig:kappaemu} shows the density dependence of the full
electron-muon thermal conductivity $\kappa_{e\mu}$ for the
selected EOSs. The thick lines give
$\kappa_{e\mu}$ at $T=10^8$ K, a typical internal temperature of a
middle-aged neutron star ($t \sim 10^4-10^5$ yr) with no enhanced
neutrino emission generated in its core (see, e.g., Ref.\ \cite{yp04}).
The conductivity $\kappa_{e\mu}$ increases with growing $\rho$
because of the increasing amount of electrons and muons in dense
matter [see Eq.\ (\ref{E:kappa_normal})]. The conductivity shows a
noticeable kink at $\rho=\rho_\mu$ associated with the appearance
of muons. The conductivities for the APR and PAL IV EOSs are close
to each other because these two EOSs give very similar amounts of
electrons and muons (Fig.\ \ref{fig:frac}). At $\rho \gtrsim 4
\times 10^{14}$ g~cm$^{-3}$, the conductivity for the PAL II EOS
is noticeably higher 
owing to larger
fractions of electrons and muons for the PAL II EOS at high
densities.

The thin solid line in Fig.\ \ref{fig:kappaemu} displays
$\kappa_{e\mu}$ for the PAL II EOS at a higher temperature
$T=10^9$ K typical of a young neutron star
(a few months after its birth in the
absence of enhanced neutrino emission; e.g., Ref.\ \cite{yp04}).
It is very close to the conductivity at $T=10^8$ K confirming
our prediction (\ref{E:kappa_normal})
that $\kappa_{e\mu}$ is almost independent
of temperature in nonsuperconducting matter because of the
importance of the Landau damping.
The Landau damping owing to the exchange of transverse
plasmons is the dominant mechanism of electron and muon
scattering; it fully regulates
$\kappa_{e\mu}$
in nonsuperconducting neutron star cores (Sec.\ \ref{sect-coll-freq}).

\begin{figure}
\begin{center}
\leavevmode
\epsfxsize=80mm
\epsfbox[40 175 565 680]{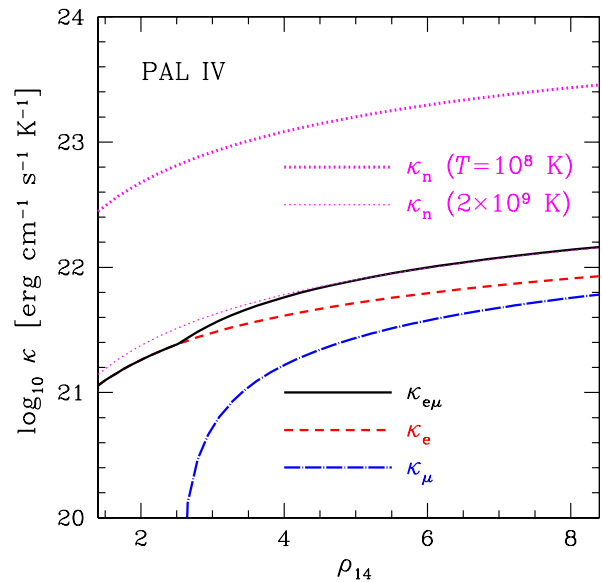}
\end{center}
\caption{(Color online) Electron (dashed line), muon (dot-dashed line),
     electron-muon (solid line) thermal conductivities 
     (calculated for $T=10^8$~K but almost
     temperature independent) and neutron conductivity
     (for $T=10^8$ and $2 \times 10^9$ K, thick and thin
     dotted lines, respectively) 
     versus density in a normal neutron star core
     for the PAL IV EOS.}
\label{fig:kappaemun}
\end{figure}

Figure \ref{fig:kappaemun} compares the thermal conductivity
provided by different particles in a nonsuperfluid neutron star
for the PAL IV EOS. We show the density dependence of the electron
conductivity $\kappa_e$, the muon conductivity $\kappa_\mu$, and the
electron-muon conductivity $\kappa_{e\mu}$ calculated at $T=10^8$~K
with the notice that these conductivities are almost temperature
independent. We also plot the neutron
conductivity $\kappa_n$ calculated
using the results of Baiko, Haensel and Yakovlev \cite{bhy01}
for $T=10^8$~K (the thick dotted line) and $2\times 10^9$~K
(the thin dotted line). It
is determined by collisions of neutrons with neutrons and protons
via nuclear interactions and demonstrates a traditional
Fermi liquid behavior ($\kappa_n \propto T^{-1}$). 
The effective masses of neutrons and
protons are set equal to 0.7 of their bare masses, and the
in-medium effects on the squared matrix elements of
nucleon-nucleon scattering are neglected. It is seen that the
neutron contribution dominates throughout the neutron star core at
$T=10^8$~K (contrary to the previous results
\cite{fi79,gy95}). However, we have
$\kappa_{e\mu}\gtrsim \kappa_n$ in the
core of a hot neutron star with $T \gtrsim 2 \times 10^9$~K.

\subsection{Thermal conductivity in superconducting matter}
\label{sect-kappa-sf}

\begin{figure}
\begin{center}
\leavevmode
\epsfxsize=80mm
\epsfbox[40 175 565 680]{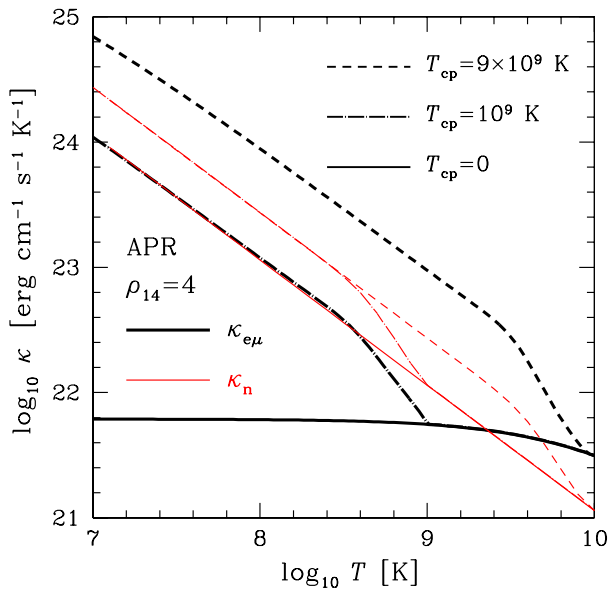}
\end{center}
\caption{(Color online) Temperature dependence of the electron-muon thermal
         conductivity $\kappa_{e\mu}$ (thick lines)
	 and the neutron conductivity $\kappa_n$ (thin lines)
	 in a neutron star core with the APR EOS
      at 
     $\rho=4 \times 10^{14}$ g~cm$^{-3}$.
     Solid lines refer to nonsuperconducting matter, while
     dot-dashed and dashed lines are for matter with
     proton superconductivity (with $T_{cp}=10^9$ and $9 \times 10^9$~K,
     respectively).    }
\label{fig-sf}
\end{figure}

Let us illustrate the effects of proton superconductivity 
on the
electron-muon thermal conductivity $\kappa_{e\mu}$. Figure
\ref{fig-sf} shows $\kappa_{e\mu}$ and $\kappa_n$ 
versus temperature
for the APR EOS at $\rho=4 \times 10^{14}$ g~cm$^{-3}$
assuming either $T_{cp}=0$ (normal matter; solid lines), or
$T_{cp}=10^9$~K (dot-dashed lines), or $T_{cp}=9\times10^9$~K
(dashed lines). The neutron conductivity is calculated using the
results of Ref.\ \cite{bhy01} under the assumption 
that neutrons are normal.

In normal matter 
$\kappa_n$ dominates over $\kappa_{e\mu}$
at $T \lesssim 2 \times 10^9$~K in agreement with 
the results of Sec.\ \ref{sect-nonsuperfluid}. 
We see that 
$\kappa_{e\mu}$ is almost temperature independent because of the
dominant contribution of the Landau damping
[Eq.\ (\ref{E:kappa_normal})].  Proton superconductivity
reduces the electron and muon collision frequencies and increases
$\kappa_{e\mu}$ as discussed in Sec.\
\ref{sect-superfluid}; 
it also slightly increases $\kappa_n$ \cite{bhy01}.
Therefore, when the temperature drops below
$T_{cp}$ and proton superconductivity sets in, $\kappa_{e\mu}$ starts
to grow up much quicker than $\kappa_n$ and becomes
comparable to or larger than $\kappa_n$. At $T
\lesssim T_{cp}/3$ this increase of $\kappa_{e\mu}$ is well
described by the asymptotic expression (\ref{restoration}),
$\kappa_{e\mu}\propto T^{-1}$. Proton superconductivity
formally restores the
Fermi liquid behavior of $\kappa_{e\mu}$ 
(Sec.\ \ref{sect-superfluid}). However, this $\kappa_{e\mu}$ is
smaller than the electron-muon conductivity
which is  calculated in Refs.\ \cite{fi79,gy95}.
Nevertheless, because $\kappa_{e\mu}$ in
superconducting matter is directly proportional to the proton
gap $\Delta \propto T_{cp}$, the enhancement of $\kappa_{e\mu}$ by
proton superconductivity makes 
$\kappa_{e\mu}$ quite large.
According to Fig.\ \ref{fig-sf},
for $T_{cp}=10^9$~K and $T \lesssim 3 \times 10^8$~K 
we have $\kappa_{e\mu}$ much larger than in normal
matter although a factor of $\sim 3$ lower than
$\kappa_n$. For $T_{cp}=3 \times 10^9$~K (not shown in the
figure) we would have $\kappa_{e\mu} \sim \kappa_n$ at any $T$
displayed.
For a stronger superconductivity
with $T_{cp}=9\times 10^9$~K, 
as seen from Fig.\ \ref{fig-sf}, $\kappa_{e\mu}$
dominates over $\kappa_n$ at any $T$.
Therefore, we obtain $\kappa_{e\mu}
\gtrsim \kappa_n$ for $T \gtrsim 2 \times 10^9$~K
in normal matter and for any $T$ in superconducting matter
with $T_{cp} \gtrsim 3 \times 10^9$~K. 

\section{Conclusions}
\label{sect-conclusions}

We have calculated the thermal conductivity of
electrons and muons $\kappa_{e\mu}$ in the cores
of neutron stars composed of neutrons, protons, electrons, and
muons taking into account possible superconductivity of
protons. This thermal conductivity is determined by
electromagnetic interactions of electrons and muons with all charged
particles. Our results are presented in the form of simple
analytic expressions and fitting formulae valid for
any equation of state of dense matter in a neutron star core
and for any ratio $T/T_{cp}$ between the temperature $T$
and proton critical temperature $T_{cp}$ in
superconducting matter ($T<T_{cp}$). A generalization of our
results to hyperonic matter (including the case
of superfluid hyperons) is straightforward (analogous to that
in Ref.\ \cite{gy95}).

Our main conclusions are as follows.

\begin{enumerate}

\item The main contribution to $\kappa_{e\mu}$ comes from
electromagnetic interactions of charged particles via the exchange of
transverse plasmons. This contribution has been neglected
in all previous calculations of $\kappa_{e\mu}$ in neutron
star cores (although it has been included in the calculation
of the thermal conductivity of quarks in quark matter \cite{hp93}
and the thermal conductivity of electrons in neutron star
crusts \cite{sy06}).

\item For normal (nonsuperconducting) protons,
$\kappa_{e\mu}$ is determined by electromagnetic interactions
of electrons and muons with all charged particles ($e$, $\mu$, $p$)
via the Landau damping of transverse plasmons [Eq.\ (\ref{E:kappa_normal})].
This thermal conductivity is temperature independent
(contrary to the traditional Fermi liquid behavior $\kappa \propto T^{-1}$).

\item The conductivity $\kappa_{e\mu}$ is mainly determined by the
symmetry energy of dense matter (rather than by the stiffness of
the equation of state) and increases with the growth of the
symmetry energy. 

\item At temperatures $T\sim 10^8$~K, the conductivity $\kappa_{e\mu}$
in a normal neutron star core is smaller than the conductivity
of neutrons $\kappa_n$, but at $T \gtrsim 2 \times 10^9$~K we obtain
$\kappa_{e\mu} \gtrsim \kappa_n$.

\item After the onset of proton superconductivity (at $T<T_{cp}$)
$\kappa_{e\mu}$ remains to be determined by electromagnetic interactions
of charged particles via the exchange of transverse plasmons but
the character of these interactions is different. First, strong
proton superconductivity greatly suppresses the collisions of
electrons and muons with protons. Second, this superconductivity
modifies the exchange of transverse plasmons (from the dissipative
Landau damping to the nondissipative regime). As a result, strong
superconductivity ($T \lesssim T_{cp}/3$) increases $\kappa_{e\mu}$,
restores its Fermi liquid behavior, and enforces $\kappa_{e\mu}$ to scale as
$\kappa_{e\mu} \propto T_{cp}/T$. For $T_{cp} \gtrsim 3 \times 10^9$~K,
we obtain $\kappa_{e\mu} \gtrsim \kappa_n$ at any $T$.

\end{enumerate}

Our results are based on previous calculations of Gnedin and
Yakovlev \cite{gy95}. We have extended them by
properly including the 
charged 
particle interactions via the
exchange of transverse plasmons and by performing exact
calculation of the rate of electron/muon scattering by proton
quasiparticles in the presence of proton superconductivity.

We expect that our results give reliable values of $\kappa_{e\mu}$
in neutron star cores. In the absence of proton superconductivity,
calculation of $\kappa_{e\mu}$ constitutes a well defined
problem, which we solve explicitly (under the only one
and well justified 
assumption of weak plasma screening). Because the solution 
involves only electromagnetic
interactions, it is universal.
It does not depend on a strong interaction model
of dense matter
and, hence, on a specific equation of state in a neutron star core.
In the presence of proton superconductivity, we use the polarization
functions of a proton gas derived in the BCS framework. We argue 
(Sec.\ \ref{dielectricfun}) that it can be a good approximation.
Even if future studies of realistic nucleon Fermi liquid
give more elaborated polarization functions, our solution
will serve as a useful basic approximation. 

In any case our results do not
solve the thermal conduction problem in neutron star cores.
The main problem is posed by the conductivity of neutrons and
protons (and other baryons if available) mediated by strong
(nuclear) interactions, especially in the presence of baryon
superfluidity. The conductivities of neutrons and protons,
$\kappa_n$ and $\kappa_p$, in normal matter have been
estimated by Flowers and Itoh \cite{fi79,fi81} (with the natural
result that $\kappa_p$ is much smaller than $\kappa_n$ because of
smaller amount of protons in neutron star cores). More accurate
estimations of $\kappa_n$ were done later by Baiko {\it et al.}
\cite{bhy01} who estimated also the
diffusive thermal conductivity $\kappa_n$ in the presence of
neutron and proton superfluidity. However, the appearance of
neutron superfluidity can trigger a specific and efficient nondiffusive
heat conduction via convective counterflow. This effect is well
known from laboratory experiments with superfluid $^4$He (e.g.,
Ref.\ \cite{tt90}); it is so efficient that immediately dissolves
any temperature gradients in superfluid $^4$He. 
Analogous effects in superfluid neutron star cores
have been mentioned in the astrophysical literature (e.g., Ref.\
\cite{fi79}) but have not been explored. Their careful examination
would be desirable.

It would also be important to study the effects of strong
magnetic fields which can greatly modify thermal conductivity
in neutron star cores. Strong Larmor rotation of charged
particles (electrons, muons, protons) about magnetic field
lines can greatly suppress thermal conduction across the
magnetic field. Similar effects in electric conductivity
have been studied in a number of works (see, e.g.,
\cite{yakovlev93} and references therein) but they are
almost unexplored for the thermal conductivity.  The magnetic
field modifies plasmon modes in dense matter and
affects plasma polarization properties, particularly, the
Landau damping. The magnetic field effects can be especially
complicated in the presence of baryon superfluidity.

All in all, further serious efforts are required to solve
the heat conduction problem in neutron star cores.
First of all, its solution is needed to model cooling
of neutron stars \cite{yp04,pgw06}
(especially in the first 100 years of
their life \cite{lattimeretal94,gyp01}),
thermal states of neutron stars in soft X-ray
transients \cite{pgw06,bbr98,hz90,hz03}
and thermal relaxation of pulsars after glitches \cite{ll02}.
We hope that our results give
a reliable contribution to the thermal conduction problem,
and we expect to study other effects of the Landau damping
on kinetic properties of neutron star cores (particularly,
on the shear viscosity) in subsequent
publications.

\begin{acknowledgments}
We are extremely grateful to M.~E.\ Gusakov
for many useful discussions and critical remarks.
This work was partly supported by
the Dynasty Foundation,
by the Russian Foundation for Basic Research
(grants 05-02-16245, 05-02-22003), and
by the Federal Agency for Science and Innovations
(grant NSh 9879.2006.2).
\end{acknowledgments}


\begin{thebibliography}{99}

\bibitem{st83}
S.~L.\ Shapiro and S.~A.\ Teukolsky, {\em Black Holes, White Dwarfs,
and Neutron Stars} (Wiley-Interscience, New York, 1983).

\bibitem{lp01}
J.~M.\ Lattimer and M.\ Prakash, Astrophys.\ J.\ {\bf 550}, 426
(2001).

\bibitem{haensel03}
P.~Haensel, in {\em Final Stages of Stellar Evolution},
edited by C.\ Motch and J.-M. Hameury (EAS Publ.\ Ser., EDP Sci., 2003)
p.\ 249.

\bibitem{lp04}
J.~M.\ Lattimer and M.\ Prakash, Science {\bf 304}, 536 (2004).

\bibitem{yp04}
D.~G.\ Yakovlev and C.~J.\ Pethick,
Annu.\ Rev.\ Astron.\ Astrophys.\
{\bf 42}, 169 (2004).

\bibitem{pgw06}
D.\ Page, U.\ Geppert, and F.\ Weber, Nucl.\ Phys.\ 
A {\bf 777}, 497 (2006).

\bibitem{bbr98}
Brown~E.~F., Bildsten~L., and Rutledge~R.~E.,  Astrophys. J. Lett.
{\bf 504}, L95 (1998).

\bibitem{hz90}
Haensel~P. and Zdunik~J.~L., Astron. Astrophys. {\bf 227}, 431
(1990).

\bibitem{hz03}
Haensel~P. and Zdunik~J.~L., Astron. Astrophys. {\bf 404}, L33
(2003).

\bibitem{lattimeretal94}
J.~M.\ Lattimer, K.~A.\ Van Riper, M.\ Prakash, and M.\ Prakash,
Astrophys.\ J.\ {\bf 425}, 802 (1994).

\bibitem{gyp01}
O.~Y.\ Gnedin, D.~G.\ Yakovlev, and A.~Y.\ Potekhin, Mon.\ Not.\
R.\ Astron.\ Soc.\ {\bf 324}, 725 (2001).

\bibitem{ll02}
M.~B.\ Larson and B.~Link, Mon.\ Not.\ R.\ Astron.\ Soc.\ {\bf 333}, 613
(2002).

\bibitem{fi79}
E.\ Flowers and N.\ Itoh,  Astrophys.\ J.\  {\bf 230}, 847 (1979).

\bibitem{fi81}
E.\ Flowers and N.\ Itoh, Astrophys.\ J.\  {\bf 250}, 750 (1981).

\bibitem{bhy01}
D.~A.\ Baiko, P.\ Haensel, and D.~G.\ Yakovlev, Astron.\
Astrophys.\ {\bf 374}, 151 (2001).

\bibitem{fi76}
E.\ Flowers and N.\ Itoh, Astrophys.\ J.\  {\bf 206}, 218 (1976).

\bibitem{gy95}
O.~Y.\ Gnedin and D.~G.\ Yakovlev, Nucl.\ Phys.\ A {\bf 582}, 697 (1995).


\bibitem{hp93}
H.\ Heiselberg and C.~J.\ Pethick,  Phys.\ Rev.\ D {\bf 48}, 2916 (1993).


\bibitem{sy06}
P.~S.\ Shternin and D.~G.\ Yakovlev, Phys.\ Rev.\ D {\bf 74}, 043004
(2006).

\bibitem{jaikumaretal05}
P.\ Jaikumar, C.\ Gale, and D.\ Page, Phys.\ Rev.\ D {\bf 72}, 123004 (2005).

\bibitem{QED}
V.~B.\ Berestetski{\u\i}, E.~M.\ Lifshitz, and L.~P.\ Pitaevskii,
\textit{Quantum Electrodynamics}
(Butterworth-Heinemann, Oxford, 1982).

\bibitem{abr84}
A.~F.\ Alexandrov, L.~S.\ Bogdankevich, and A.~A.\ Rukhadze,  {\it
Principles of Plasma Electrodynamics}  (Springer-Verlag, Berlin,
Heidelberg, New York, Tokyo, 1984), Springer Series in
Electrophysics, Vol.\ 9.

\bibitem{lp80}
E.~M.\ Lifshitz and L.~P.\ Pitaevski{\u\i},
\textit{Statistical Physics, Part~2}
(Pergamon, Oxford, 1980). 


\bibitem{sybr70}
J. Sykes and G.~A. Brooker, Ann. Phys. (NY) {\bf 56}, 1 (1970).

\bibitem{and87}
R.~H.\ Anderson, C.~J.\ Pethick and K.~F.\ Quader, Phys.\ Rev.\ B {\bf
35},  1620 (1987).

\bibitem{Baym91}
G.\ Baym and C.~J.\ Pethick, \textit{Landau Fermi-Liquid Theory.
Concepts and Applications} (Wiley, New-York, 1991).


\bibitem{ls01}
U.\ Lombardo and H.-J.\ Schulze,
in \textit{Physics of Neutron Star Interiors},
edited by D.\ Blaschke, N.\ Glendenning, and A.\ Sedrakian
(Springer, Berlin, 2001), p.\ 30. 

\bibitem{ly94}
K.~P.\ Levenfish and D.~G.\ Yakovlev,
Astron.\ Rep.\
{\bf 38}, 247 (1994).


\bibitem{alf06}
P.~I. Arseev, S.~O. Loiko, and N.~K. Fedorov, Usp. Fiz. Nauk {\bf
49}, 1 (2006) [Phys. --- Usp. {\bf 49}, 1 (2006)].

\bibitem{kr04}
J.~Kundu and S.~Reddy, Phys.\ Rev.\ C {\bf 70}, 055803 (2004).

\bibitem{Mattis58}
D.~C.\ Mattis and J.\ Bardeen, Phys.\ Rev.\ {\bf 111}, 412 (1958).

\bibitem{Abrikosov1975}
A.~A.\ Abrikosov, L.~P.\ Gorkov, and I.~E.\ Dzyaloshinski,
\textit{Methods of Quantum
Field Theory in Statistical Physics} (Courier Dover Publications,
1975).


\bibitem{apr98}
A.\ Akmal, V.~R.\ Pandharipande, and D.~G.\ Ravenhall,
Phys.\ Rev.\ C {\bf 58}, 1804 (1998). 

\bibitem{pal88}
M.\ Prakash, T.~L.\ Ainsworth, and J.~M.\ Lattimer,
Phys.\ Rev.\ Lett.\  {\bf 61}, 2518 (1988). 

\bibitem{pa92}
D.~Page and J.~H.\ Applegate, Astrophys.\ J.\ Lett.\ {\bf 394},
L17 (1992).



\bibitem{tt90}
D.~R.~Tilley and J.\ Tilley,
\textit{Superfluidity and
Superconductivity} (IOP Publishing, Bristol, 1990).

\bibitem{yakovlev93}
D.~G. Yakovlev, 
in \textit{Strongly Coupled Plasma Physics}, edited by H.~M.\ Van
Horn and S.\ Ichimaru
(University of Rochester, Rochester, 1993), p.\ 157. 





\end{thebibliography}
\end{document}